\documentclass[transmag]{IEEEtran}

\usepackage[T1]{fontenc}

\usepackage{graphicx}
\usepackage{caption}
\usepackage{subcaption}
\usepackage{color}

\IEEEoverridecommandlockouts
\usepackage{mathtools}
\usepackage{amsmath}
\usepackage{algorithmic,algorithm}
\usepackage{cite}
\usepackage{tabularx}
\usepackage{amsfonts}

\allowdisplaybreaks

\usepackage{graphicx}

\usepackage{amsfonts,amssymb,amsmath}
\usepackage{amsthm}

\newtheorem{theorem}{Theorem}

\newtheorem{prop}{Proposition}
\DeclareMathOperator*{\argmax}{argmax}

\newtheoremstyle{lemma}
  {0}
  {0}
  {}
  {}
  {\itshape}
  {:}
  {.5em}
  {\thmname{#1}\thmnumber{ #2}\thmnote{ (#3)}}
\theoremstyle{lemma}
\newtheorem{lemma}{Lemma}
\newcommand{\algorithmicbreak}{\textbf{break}}
\newcommand{\BREAK}{\STATE \algorithmicbreak}

\def\BibTeX{{\rm B\kern-.05em{\sc i\kern-.025em b}\kern-.08em T\kern-.1667em\lower.7ex\hbox{E}\kern-.125emX}}
\begin{document}

\title{QoS Aware Robot Trajectory Optimization with IRS-Assisted Millimeter-Wave Communications}

\author{Cristian~Tatino,~\IEEEmembership{Member,~IEEE,}
	Nikolaos~Pappas,~\IEEEmembership{Senior Member,~IEEE,}
	and~Di~Yuan,~\IEEEmembership{Senior Member,~IEEE}
	\thanks{This work extends the preliminary study in~\cite{OurRobot} and was supported in part by CENIIT, ELLIIT, and by the European Union's Horizon 2020 research and innovation programme under the Marie Sklodowska-Curie grant agreement No. 643002 (ACT5G).
		
		Cristian~Tatino was with Department of Science and Technology (ITN), Link\"{o}ping University,
		Sweden. He is now with Ericsson AB, 16483 Stockholm, Sweden (Email: cristian.tatino@ericsson.com).
		Nikolaos~Pappas and Di~Yuan are with Department of Science and Technology (ITN), Link\"{o}ping University,
		Sweden (Email: nikolaos.pappas@liu.se, di.yuan@liu.se}}

\IEEEtitleabstractindextext{\begin{abstract}
		
In this paper, we consider the motion energy minimization problem for a robot that uses millimeter-wave (mm-wave) communications assisted by an intelligent reflective surface (IRS). The robot must perform tasks within given deadlines and it is subject to uplink quality of service (QoS) constraints. This problem is crucial for fully automated factories that are governed by the binomial of autonomous robots and new generations of mobile communications, i.e., 5G and 6G. In this new context, robot energy efficiency and communication reliability remain fundamental problems that couple in optimizing robot trajectory and communication QoS. More precisely, to account for the mutual dependency between robot position and communication QoS, robot trajectory and beamforming at the IRS and access point all need to be optimized. We present a solution that can decouple the two problems by exploiting mm-wave channel characteristics. Then, a closed-form solution is obtained for the beamforming optimization problem, whereas the trajectory is optimized by a novel successive-convex optimization-based algorithm that can deal with abrupt line-of-sight (LOS) to non-line-of-sight (NLOS) transitions. Specifically, the algorithm uses a radio map to avoid collisions with obstacles and poorly covered areas. We prove that the algorithm can converge to a solution satisfying the Karush-Kuhn-Tucker conditions. The simulation results show a fast convergence rate of the algorithm and a dramatic reduction of the motion energy consumption with respect to methods that aim to find maximum-rate trajectories. Moreover, we show that the use of passive IRSs represents a powerful solution to improve the radio coverage and motion energy efficiency of robots.\end{abstract}
	
}

\maketitle




%
\IEEEpeerreviewmaketitle

\section{Introduction}
\label{sec:Intro}

Robotic and wireless technologies are driving the new industrial revolution, i.e., Industry 4.0, and playing a crucial role in the digital transition of manufacturing processes, warehousing, and logistics~\cite{Ericsson1}. However, the massive exploitation of robots and the rising of new industrial applications, with stringent quality of service (QoS) requirements, will stress the performance of the next generation of mobile communications, i.e., 6G. Specifically, real-time industrial applications, such as augmented and virtual reality for assisted manufacturing or mining, may require Gbps for peak data rates~\cite{6G,Cheffena}. Moreover, swarms consisting of hundreds of sensing robots in the warehouses of the future may need to operate with latency and reliability requirements of $1$~ms and up to $99.9999$\%, respectively~\cite{6G,Swarm}. 

Millimeter-wave (mm-wave) spectrum has been identified as a possible solution for wireless communications in industrial scenarios~\cite{Ericsson3}. However, high-band communications suffer from high blockage sensitivity~\cite{Block,OurBlock} that reduces communication reliability when a robot moves in environments with obstacles. In addition to avoiding obstacles, trajectory planning highly affects mm-wave performance as it determines whether the robot is in line-of-sight (LOS) or non-line-of-sight (NLOS). Moreover, robots are battery-limited and have tasks that are usually characterized by stringent deadlines. By optimizing the robot's movement, it is possible to dramatically decrease the robot energy consumption with a significant reduction in the total electrical energy consumption for manufacturing processes. Consequently, in the last decades, robot trajectory planning has been one of the most relevant problem in robotics \cite{En1,RobEn1,RobEn2,MEn,RobReview1,RobReview2} and it has assumed particular importance for wirelessly connected robots, where trajectory must be optimized according to the radio coverage \cite{CoCP,RobEn3,RobRel}. 

Beside trajectory optimization, several solutions have been proposed to enhance coverage in mm-waves scenarios, e.g., relays~\cite{Our2} and intelligent reflective surfaces (IRSs)~\cite{IRSmmwave2,IRSmmwave3}. The latter consist of arrays of reflective elements that can be electronically controlled to adjust the angle and the phase of the reflected signals to be either added coherently or destructively for the receiver~\cite{IRS_Renz,Liaskos}. Due to the short wavelength at mm-wave frequency ranges, IRSs with many reflective elements can be deployed to improve throughput and reliability of robot communications. Specifically, IRSs can provide alternative signal paths when the LOS path is blocked. Moreover, in comparison to active relays, the negligible energy consumption and the lower cost~\cite{IRSvsRelay,IRS_En} make passive IRSs ideal candidates for increasing the energy efficiency of fully autonomous robots. However, beamforming at the IRS must be set according to the channel that depends on the robot trajectory. Therefore, we consider a trajectory and beamforming co-optimization to minimize the motion energy consumption of a wirelessly connected robot in IRS-assisted mm-wave scenarios. To solve this problem, we propose a modified successive convex optimization (SCO) algorithm that accounts for the knowledge of the environment and a radio map to avoid collisions and satisfy time and QoS constraints.

\subsection{Related Works}
\label{sec:rel}

Energy-aware trajectory optimization has been one of the most crucial problems in robotics~\cite{En1,RobEn1,RobEn2,MEn}. In~\cite{En1}, the authors model the power consumption of a DC motor-equipped robot as a function of the speed. By optimally controlling the robot's speed, it is possible to achieve up to $50$\% energy-saving. The work in~\cite{RobEn1} uses a graph-based method and A* algorithm to determine the robot's minimum cost path, where the cost of an edge represents the corresponding motion energy consumption. A convex optimization approach is adopted in~\cite{RobEn2}, which presents an alternating quadratic programming method to determine the path that minimizes the energy consumption in scenarios with obstacles. Multi-robot scenarios are studied in~\cite{MEn}, where the authors propose a distributed algorithm for optimizing locations and times of robots' rendezvous. This problem includes the battery level of the robots as a constraint. The corresponding energy consumption is derived from the optimal control problem that minimizes the energy consumption along the robot trajectory. In contrast to the previously mentioned studies, the work in~\cite{CoCP} deals with wirelessly connected robots. More precisely, the authors propose joint robot communication and motion energy minimization by controlling the transmit power and the robot's speed along a fixed trajectory. 

The possibility to control the robot motion introduces a new degree of freedom for resource allocation problems in wireless communications. In the past few years, several studies have explored this topic~\cite{CoCP,RobEn3,RobRel,MPC3,MCRA,MCRA2,MCRA3}. Similar to~\cite{CoCP}, the work in~\cite{RobEn3} proposes an optimal control problem for motion and communication energy minimization subject to a certain amount of data to be transmitted and power limit. In this case, multiple robots transmit data to an access point (AP) in a scenario without obstacles. In~\cite{RobRel}, the authors define a convex optimization problem to minimize the energy consumption of a moving relay and multiple mobile sensing robots by controlling trajectories and transmit power. The sensing robots move in a scenario without obstacles along paths that depend on the assigned tasks. Joint task and trajectory optimization for multiple robots is explored in~\cite{MPC3,MCRA,MCRA2,MCRA3}, where~\cite{MCRA3} deals with directional communication scenarios. 

Directional communications are typical for mm-wave communications and the latter has recently attracted the interest of industrial applications~\cite{Cheffena,Ericsson2,mmInd,mmInd2,MAPP}. Mm-wave connected robots are considered in~\cite{MAPP}, where the authors present several instances of an association and path planning problem in multi-AP mm-wave networks. A graph-based algorithm is used in~\cite{MAPP} to minimize handovers and travel time of multiple robots, where a radio map is used to account for obstacles and communication blockages. However, the potentials of mm-waves transmissions for wirelessly connected robots need to be further explored. Several studies have been performed for unmanned aerial vehicle (UAV) scenarios, as shown in~\cite{UAVmmWave2}, where joint communication and trajectory planning problems are particularly relevant~\cite{UAV1,UAV2,Mano,UAV3,UAV4,RadioMap}. Specifically, similar to~\cite{MAPP}, the work in~\cite{RadioMap} uses a graph-based method and a radio map to optimize UAVs' flying distance while ensuring a target QoS requirement. 

In the last few years, researchers have considered the application of IRSs to enhance UAV-user communications~\cite{UAVIRS1,UAVIRS2,UAVIRS3}. Specifically, in~\cite{UAVIRS1} and~\cite{UAVIRS2}, authors study UAV-users communications assisted by stationary IRSs. They study joint UAV trajectory and IRS beamforming optimization problems for maximizing the received power at the users. A similar scenario is considered in~\cite{UAVIRS3}, where IRSs mounted on UAVs are used to establish indirect LOS links to maximize the minimum rate among user clusters. The resulting joint UAV positioning and IRS beamforming optimization problem is solved by using a hybrid particle swarm optimization-based heuristic algorithm. IRSs for enhancing low-frequency robot-AP communications are studied in~\cite{Mu_2021}, where the authors use a graph-based method to minimize the robot traversal time while ensuring a minimum data rate requirement at each location along the trajectory. However, as will be detailed in later sections, maximum rate trajectories and graph-based methods would provide high energy consumption or infeasible trajectories if applied to our problem.

\subsection{Contributions}
\label{sec:rel}
The contributions of this paper are summarized as follows:
\begin{itemize}

\item We consider a novel robot trajectory optimization problem with IRS-assisted mm-wave communications. The problem aims to minimize the motion energy consumption while satisfying minimum average data rate and deadline constraints. Moreover, the robot must avoid collisions with obstacles. To solve this problem, we account for the mutual dependence of the energy consumption and achieved data rate on the robot trajectory and beamforming at the IRS and AP. To the best of our knowledge, energy-efficient trajectory planning problems have not been considered for wirelessly connected robots using mm-waves, which have peculiar signal propagation conditions.


\item In mm-wave scenarios, when obstacles create abrupt LOS-NLOS transitions, the received data rate is not a convex function of the robot's position. Therefore, to solve the problem mentioned above, we propose a novel SCO-based algorithm, which exploits a radio map to establish if a trajectory solution satisfies the QoS constraint. Previous SCO-based algorithms for UAVs, e.g., in~\cite{UAVIRS1,UAVIRS2}, may lead to infeasible paths or higher energy consumption if applied to our problem. Moreover, UAV communications do not present the same characteristics as wirelessly connected robots of which the altitude cannot be adapted. As in~\cite{Mu_2021}, graph-based methods can also include radio maps, but they may perform poorly when applied to our problem and cannot account for the average data rate constraint.


\item We prove that the proposed SCO algorithm converges and, under certain conditions, it converges to a point satisfying KKT conditions. The presented method can find trajectories that avoid collisions and satisfy the QoS requirement by using radio map information. The proposed algorithm can dramatically reduce the robot's energy consumption with respect to trajectories that maximize the data rate. Finally, we show that IRSs can enhance the motion energy efficiency for QoS-constrained wirelessly connected robots. Specifically, by increasing the number of reflective elements of the IRS, the solution of the algorithm converges to the trajectory corresponding to the minimum energy consumption.
\end{itemize}


The rest of the paper is organized as follows. In Section~\ref{sec:Ass} we describe the system model. In Section~\ref{sec:Prob}, we formulate the problem; moreover, we decouple beamforming and trajectory optimizations. In Section~\ref{sec:Traj_Opt}, we solve the latter by using an SCO algorithm, and in Section~\ref{sec:Res}, we provide performance evaluation. Finally, Section~\ref{sec:Conc} concludes the paper.

\section{System Model and Assumptions}
\label{sec:Ass}
We consider an industrial scenario, e.g., an industrial plant, where a robot moves from a starting position $q_{s}$ to its goal $q_{d}$ within a time horizon of fixed duration. The robot moves on the horizontal plane of a 3D restricted area containing several 3D obstacles. These are represented by a set $\mathcal{O}$ of cylinders with elliptic bases and given heights.\footnote{Note that arbitrarily shaped obstacle can be often approximated by the intersection and the union of several convex shapes~\cite{Borrelli}. In this paper, we consider 3D cylinders with elliptic bases.} The area is covered by an AP using mm-waves to which the robot needs to transmit uplink data by maintaining a given communication QoS. This is expressed as a minimum average data rate requirement\footnote{A minimum average data rate requirement can model applications where robots can store data in a buffer and transmit when channel conditions are favorable. Moreover, this choice makes the problem more general and the solution presented in this manuscript can be easily extended to instantaneous data rate requirements.} ($r_{min}$). The robot is equipped with a single antenna, whereas, the AP is equipped with $N$ antennas. The robot-AP communication is assisted by an IRS consisting of a uniform linear array (ULA) of $M$ reflective elements, of which the phase shifters are adjusted by a controller, which shares the channel state information (CSI) with the AP. A scheduler, which we assume to be co-located with the controller, optimizes the robot trajectory. The goal is to minimize the motion energy consumption accounting for both active and passive beamforming at the AP and IRS, respectively.


\textit{Notations}: $(.)^{T}$ and $(.)^{H}$, represents the transpose and the conjugate transpose, respectively; $diag(.)$ returns the diagonalization of a vector and $arg(.)$ denotes the phase of a complex number. Finally, $\lVert . \lVert_{n}$ represents the $n$-norm. Moreover, a summary of the notation is available in Table~\ref{par}.
 
\begin{figure}[tb]
	\centering
	\includegraphics[width=8.5cm]{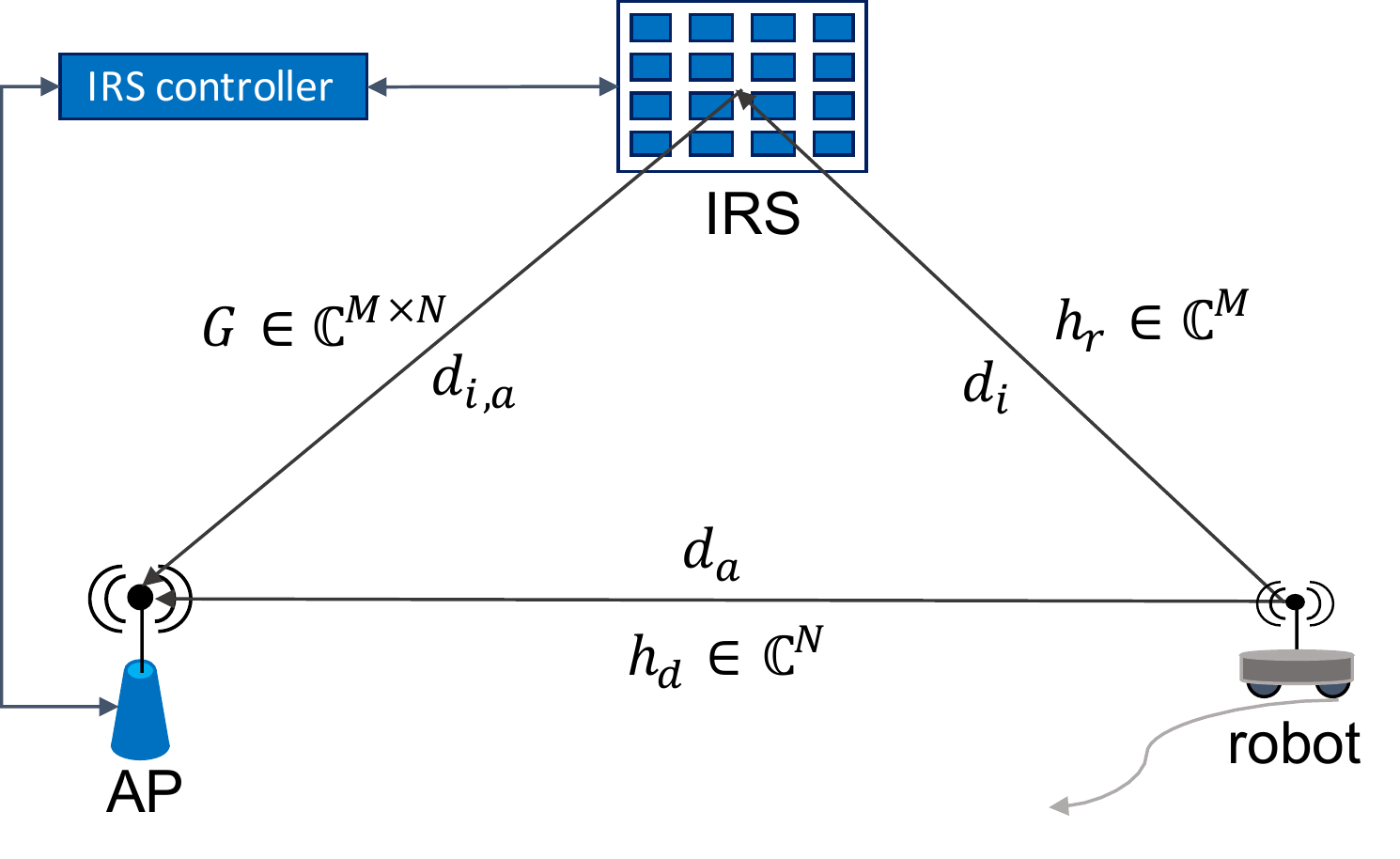}
	\caption[]{A scenario consisting of an IRS-aided robot uplink communication.}
	\label{fig:Scen}
\end{figure}

\subsection{Robot Motion Model}
\label{sec:Channel}
The robot must avoid collisions with obstacles and reach the destination within a given deadline. We divide the time horizon, which is defined by the deadline, in $K$ small slots of duration $\Delta_{t}$. Within a timeslot, the robot can travel for a maximum distance of $D_{max}=v_{max}\Delta_{t}$, where, $v_{max}$ is the maximum speed. A trajectory of the robot is represented as a sequence of $K+1$ positions, i.e., $\boldsymbol{q}=[q_{0},q_{1},...,q_{K}]$, where $q_{0}=q_{s}$ and $q_{K}=q_{d}$. The terms  $q_{k}=[x_{k},y_{k}], \; k=0,...,K,$ represent the Cartesian coordinates of the robot positions on the horizontal plane along the trajectory.

Let $q_{a}=[x_{a},y_{a}]$ and $q_{i}=[x_{i},y_{i}]$ represent the fixed positions of the AP and the IRS, respectively. The altitude of the robot is fixed at its antenna height $z_{r}$, whereas $z_{a}$ represents the height at which the AP is installed and $z_{i}$ the height of the IRS. Let $v_{k}$ be the speed of the robot at the $k$-th timeslot. Then, the motion energy consumption of the DC motor-equipped robot along the path can be written as follows~\cite{CoCP}: 
\begin{align*} 
E=&\sum_{k=1}^{k=K}E_{k}=\sum_{k=1}^{k=K}c_{1}v_{k}^{2}\Delta_{t}+c_{2}v_{k}\Delta_{t}+c_{3}\Delta_{t}=\\
&\sum_{k=1}^{K} c_{1} \frac{\lVert q_{k}-q_{k-1}\lVert_{2}^{2}}{\Delta_{t}}+c_{2}\lVert q_{k}-q_{k-1}\lVert_{2}+c_{3}\Delta_{t}\stepcounter{equation}\tag{\theequation}\label{eq:en_mot},
\end{align*} 
where, $E_{k}$ is the energy consumption in the $k$-th timeslot, and $c_{1}$, $c_{2}$, and $c_{3}$ are positive constants depending on the characteristics of the robot and external load.

\subsection{Channel Model}
\label{sec:Channel}
As shown in Fig.~\ref{fig:Scen}, let $h_{r}\in \mathbb{C}^{M}$ be the channel vector between the robot and the IRS and $G \in \mathbb{C}^{M\times N}$ denote the channel matrix between the IRS and the AP. The direct channel between the robot and the AP is represented by vector $h_{d}\in \mathbb{C}^{N}$. Then, the received baseband signal at the AP when the robot is at position $q_{k}$ can be written as follows:
\begin{align*} 
y_{k}&=\left(h_{r,k}^{H} \Phi_{k} G_{k}+h_{d,k}^{H}\right)w_{k}\sqrt{p_{t}}s_{k}+\eta_{k} \stepcounter{equation}\tag{\theequation}\label{eq:rec_sig},
\end{align*} 
where, $s_{k}$ and $p_{t}$ are the transmit signal and the transmit power in the uplink, respectively, $\eta_{k} \sim \mathcal{CN}(0,\sigma^{2})$ denotes the additive white Gaussian noise (AWGN). The term $w_{k} \in \mathbb{C}^{M}$ is the normalized beamforming vector at the AP, and $\Phi_{k}=\mbox{diag}\left(e^{j\theta_{1,k}},...,e^{j\theta_{M,k}}\right)$ is a diagonal matrix that accounts for the phase shifts $\theta_{m,k} \in  \left[ 0,2\pi  \right]$ associated with the reflective elements of the IRS. Due to the high path loss of mm-wave transmissions, signals that are reflected more than once are subject to severe attenuations and are not considered in~\eqref{eq:rec_sig}. Thus, the received signal-to-noise ratio (SNR) for $q_{k}$ can be written as follows:
\begin{align*} 
\mbox{SNR}_{k}=\frac{\lvert \Big(h_{r,k}^{H} \Phi_{k} G_{k}+h_{d,k}^{H}\Big)w_{k}\rvert^{2}}{\sigma^{2}}p_{t},\stepcounter{equation}\tag{\theequation} \label{eq:snr}
\end{align*} 
where, the superscript $H$ represents the hermitian. Moreover, let $d_{i,k}=\sqrt{(z_{r}-z_{i})^{2}+\lVert q_{k}- q_{i}\lVert^{2}_{2}}$, and $d_{a,k}=\sqrt{(z_{r}-z_{a})^{2}+\lVert q_{k}- q_{a}\lVert^{2}_{2}}$ be the robot-IRS and robot-AP distances, respectively. Then, channel vectors $h_{r,k}$ and $h_{d,k}$ can be modeled as follows:
\begin{align*} 
h_{r,k}=\sqrt{\rho d_{i,k}^{-\nu}}\tilde{h}_{r,k}\stepcounter{equation}\tag{\theequation}\label{eq:cha_r},
\end{align*} 
\begin{align*} 
h_{d,k}=\sqrt{\rho d_{a,k}^{-\mu}}\tilde{h}_{d,k}\stepcounter{equation}\tag{\theequation}\label{eq:cha_d},
\end{align*} 


where, $\tilde{h}_{r,k} \sim \mathcal{CN}(0,I)$ and $\tilde{h}_{d,k} \sim \mathcal{CN}(0,I)$ are complex gaussian vectors whose elements are independent and identically distributed (i.i.d) with zero means and unit variances. The term $\rho$ is the path loss at the reference distance of $1$ m, and $\nu$ and $\mu$ are the path loss exponents of the reflected and direct channels, respectively. 

Finally, for fixed $\Phi_{k}$, $w_{k}$, and position $q_{k}$, we obtain the data rate by using the Shannon's formula as follows:
\begin{align*} 
&\mbox{r}_{k}=\mbox{B}_{w} \mbox{log}_{2}\left(1+SNR_{k}\right)=\\
&\mbox{B}_{w} \mbox{log}_{2}\left(1+\frac{\lvert \Big(\sqrt{\rho d_{i,k}^{-\nu}}\tilde{h}_{r,k}^{H} \Phi_{k} G_{k}+\sqrt{\rho d_{a,k}^{-\mu}}\tilde{h}_{d,k}^{H}\Big)w_{k}\rvert^{2}}{\sigma^{2}}p_{t}\right).\stepcounter{equation}\tag{\theequation} \label{eq:rate_k}
\end{align*} 

To obtain~\eqref{eq:rate_k} we use~\eqref{eq:snr}, \eqref{eq:cha_r}, and~\eqref{eq:cha_d} and the term $\mbox{B}_{w}$ represents the system bandwidth. Let $\boldsymbol{\mbox{r}}=\Big[\mbox{r}_{0},\mbox{r}_{1},...,\mbox{r}_{K}\Big]$ be a vector, of which the elements represent the data rates along the robot trajectory $\boldsymbol{q}=[q_{0},q_{1},...,q_{K}]$. Thus, the average data rate for a trajectory $\boldsymbol{q}$ is given by:
\begin{align*} 
&\bar{\boldsymbol{\mbox{r}}}=\frac{1}{K}\sum_{k=0}^{K}\mbox{r}_{k}.\stepcounter{equation}\tag{\theequation} \label{eq:avg_rate}
\end{align*}

We can observe that $\bar{\boldsymbol{\mbox{r}}}$ is a function of $\Phi_{k}$, $w_{k}$, and $q_{k}$. The latter is included in $d_{i,k}$, and $d_{a,k}$. Thus, the robot's position and beamforming affect the data rate, which in turn affects the trajectory due to the QoS constraint. In the next section, we formulate the joint beamforming and trajectory optimization problem introduced in this section.

\begin{table*}[!t]
	\renewcommand{\arraystretch}{1.0}
	\caption{Summary of the notation.}
	\label{par}
	\centering
	\begin{tabular}{ l | l || l | l }
		\hline
		$\boldsymbol{q}$ & robot trajectory &  $q_{k}$ & $k$-th robot position along the trajectory\\
		$q_{s}$ & robot's starting position &$q_{d}$& robot's final position  \\
		$q_{a}$ & AP's position & $q_{i}$ & IRS's position\\
		$z_{a}$ & AP's height & $z_{i}$ & IRS's height\\
		$d_{a,k}$ & robot-AP distance when the robot is at position $q_{k}$ & $d_{i,k}$ & robot-IRS distance when the robot is at position $q_{k}$\\
		$\mathcal{O}$  & obstacle set & $q_{c,o}$ & center's position of obstacle $o\in\mathcal{O}$ \\
		$K$ & number of timeslots & $r_{min}$ & minimum average data rate requirement\\
		$D_{max}$ & maximum distance that a robot can travel in a timeslot & $N$ & number of antennas at the AP\\ 
		$M$ & number of reflective elements at the IRS & $G_{k}$ & AP-IRS channel when the robot is at position $q_{k}$\\ 
		$h_{r,k}$ & robot-IRS channel  when the robot is at position $q_{k}$& $h_{d,k}$ & robot-AP channel  when the robot is at position $q_{k}$ \\
		$\theta_{m,k}$ & phase shift of reflective element $m$ when the robot is at position $q_{k}$ & $\Phi_{k}$ & phase shifts matrix when the robot is at position $q_{k}$ \\
		$w_{k}$ & normalized beamforming vector when the robot is at position $q_{k}$ & $B_{w}$ & system bandwidth \\
		$E$ & total robot motion energy consumption & $E_{k}$ & robot motion energy consumption in timeslot $k$\\ 
		$\mbox{SNR}_{k}$ & received SNR when the robot is at position $q_{k}$ & $r_{k}$ & achieved data when the robot is at position $q_{k}$\\ 
		$\boldsymbol{\mbox{r}}$ & achieved data rate vector along the robot trajectory & $\bar{\boldsymbol{\mbox{r}}}$ & average achieved data rate\\ 
		$\widehat{\mbox{SNR}}^{*}_{k}$ &estimated optimal SNR when the robot is at position $q_{k}$ & $\bar{\boldsymbol{\mbox{r}}}^{*}$ & optimized data rate\\
		$\bar{\boldsymbol{\mbox{r}}}^{*}_{apx}$ & concave approximation of the average achieved data rate & $\bar{\boldsymbol{\mbox{r}}}_{map}$ & average achieved data rate obtained by the radio map \\
		$T_{k}$ & trust region for $q_{k}$ & $\tau$ &trust region reduction parameter\\ 
		$\widehat{\nu}$ & estimated robot-IRS path loss exponent & $\widehat{\mu}$ & estimated robot-AP path loss exponent\\ 
		$\boldsymbol{q}_{0}$ & initial solution for Algorithm 1 (RMAP) & $\boldsymbol{q}_{j}$  & solution of Algorithm 1 (RMAP) at iteration $j$ \\
		\hline
	\end{tabular}
\end{table*}

\section{Problem Formulation}
\label{sec:Prob}
In this section, we formulate the problem introduced in 
Section~\ref{sec:Ass}. Let $\boldsymbol{\Phi}=[\Phi_{0},\Phi_{1},...,\Phi_{K}]$ and $\boldsymbol{w}=[w_{0},w_{1},...,w_{K}]$, then the joint robot trajectory and beamforming problem can be formulated as follows:
\begin{subequations}
\begin{align}
        P1:&\min_{\boldsymbol{q},\boldsymbol{\Phi},\boldsymbol{w}}E\label{opt}\\ 
        \text{s.t.}& \;\bar{\boldsymbol{\mbox{r}}}\ge \mbox{r}_{min}\label{con_Rate},\\
        		&\lVert q_{k}-q_{k-1} \lVert_{2}  \le D_{max}, \; \; k=1,...,K, \label{Con_Dist}\\
		&q_{0}=q_{s}, \; \; q_{K}=q_{d},\label{Con_Traj}\\
		&\left(q_{k}-q_{c,o}\right)^{T}P_{o}^{-1}\left(q_{k}-q_{c,o}\right)\ge d_{s},  \; \forall k, \forall o\in \mathcal{O}, \label{Con_Ob}\\
		&\lVert w_{k} \lVert_{2}^{2}  \le 1, \; \; \forall k, \label{Con_w}\\
		&\Phi_{k}=\mbox{diag}\left(e^{j\theta_{1k}},...,e^{j\theta_{Mk}}\right), \; \forall k, \label{Con_Phi}\\
		&0 \le \theta_{m,k}\le 2\pi, \; \; \forall m, \; \forall k, \label{Con_theta}
\end{align}
\end{subequations}
where, the objective function~\eqref{opt} represents the total robot motion energy consumption along the trajectory given by~\eqref{eq:en_mot}. Note that the communication energy consumption of the robot is negligible with respect to the motion energy consumption; hence, the objective function does not include the former. The first constraint~\eqref{con_Rate} represents the QoS requirement to complete the task, where $\bar{\boldsymbol{\mbox{r}}}$ is defined in~\eqref{eq:avg_rate} and $\mbox{r}_{min}$ is the minimum required average data rate. Constraints~\eqref{Con_Dist} allow the robot to move in a timeslot for a maximum distance of $D_{max}$, whereas~\eqref{Con_Traj} fix the starting and the goal positions. To avoid collisions with obstacles, we include~\eqref{Con_Ob}. More precisely, as described in the previous section, obstacles are approximated by the intersection and the union of several ellipsoids $o \in \mathcal{O}$ on the horizontal plane. Each of them is described by a center $q_{c,o}$, and a symmetric and positive definite matrix $P_{o}$. The latter defines the length of the axis and the rotation of the ellipse. The term $d_{s}\ge1$ represents a safety distance between the robot and the obstacle. Finally, constraints~\eqref{Con_w} and~\eqref{Con_theta} impose the norm of $w_{k}$ to be at most one and $\theta_{m,k}$ to be continuous, respectively.
 
Problem $P1$ is non-linear and non-convex. However, as we show in the following sections, it is possible to decouple the beamforming and the trajectory optimization problems. More precisely, we maximize the left-hand side (LHS) of~\eqref{con_Rate} in $P1$ by deriving closed-forms of $\boldsymbol{\Phi}$ and $\boldsymbol{w}$ that maximize the average data rate for each trajectory. Then, we can obtain an optimization problem equivalent to $P1$, of which the only optimization variable is $\boldsymbol{q}$. This trajectory optimization problem is solved in Section~\ref{sec:Traj_Opt} by using an SCO-based algorithm.




%

\subsection{Average Rate Maximization}
\label{sec:Avg_Rate}
In this section, we first find closed-form solutions of $\boldsymbol{\Phi}$ and $\boldsymbol{w}$ that maximize average data rate $\bar{\boldsymbol{\mbox{r}}}$. Specifically, for a fixed trajectory we solve the following problem:
\begin{subequations}
\begin{align}
        P2:&\max_{\boldsymbol{\Phi},\boldsymbol{w}}\bar{\boldsymbol{\mbox{r}}}\label{opt2}\\ 
		 \text{s.t.}& \; \eqref{Con_w},\eqref{Con_Phi},\eqref{Con_theta},\nonumber
\end{align}
\end{subequations}
where, $\bar{\boldsymbol{\mbox{r}}}$ is given by~\eqref{eq:avg_rate}. 

\begin{prop}\label{prop1}
By replacing the LHS of~\eqref{con_Rate} in $P1$, with the optimum data rate resulting from solving $P2$, we obtain a trajectory optimization problem that is equivalent to $P1$.
\end{prop}
\vspace{-0.5em}
\begin{proof}
We first note that in $P1$, the LHS of~\eqref{con_Rate} is the only expression that depends on $\Phi_{k}$ and $w_{k}$, which are not contributing to the cost~\eqref{opt}. Moreover, by solving $P2$, we obtain an optimum data rate expression $(\bar{\boldsymbol{\mbox{r}}}^{*})$ that depends only on $\boldsymbol{q}$ such that $\bar{\boldsymbol{\mbox{r}}}^{*} \ge \bar{\boldsymbol{\mbox{r}}} \; \;\forall \, \boldsymbol{q}$. Thus, by replacing the LHS of~\eqref{con_Rate} in $P1$ with $\bar{\boldsymbol{\mbox{r}}}^{*}$ we obtain an optimization problem that depends only on $\boldsymbol{q}$ with a feasible region that includes the feasible region of $P1$.
\end{proof}
\vspace{-0.5em}

To solve $P2$, we can assume that the IRS and the AP are installed with a LOS link. Since in mm-wave communications the LOS path presents a much higher gain than the sum of NLOS paths, the IRS-AP channel can be approximated by a rank-one matrix~\cite{IRSmmwave3}:
\begin{align*} 
G_{k}=\sqrt{NM\rho d_{ia}^{-2}}\tilde{G}_{k}=\sqrt{NM}\gamma \tilde{a}_{k}\tilde{b}_{k}^{T}. \; \forall k,\stepcounter{equation}\tag{\theequation} \label{eq:cha_G}
\end{align*} 
where, $\gamma=\sqrt{\rho d_{ia}^{-2}}$. The term $d_{ia}$ is the distance between the AP and the IRS that is fixed and does not depend on $q_{k}$. The path loss exponent of the LOS path between the AP and the IRS is two and $\rho$ accounts for the path loss at the reference distance and antenna gain. The terms $\tilde{a}_{k} \in \mathbb{C}^{M}$ and $\tilde{b}_{k} \in \mathbb{C}^{N}$ are the normalized array response vectors in at $q_{k}$ associated with the IRS and the AP, respectively. These can be expressed as follows:
\begin{align*} 
\tilde{a}_{k}=\frac{1}{\sqrt{M}}\left[1,e^{-j\frac{2\pi}{\lambda}l\alpha_{k}},...,e^{-j\frac{2\pi}{\lambda}l(M-1)\alpha_{k}}\right],\stepcounter{equation}\tag{\theequation} \label{eq:arr_a}
\end{align*} 
\begin{align*} 
\tilde{b}_{k}=\frac{1}{\sqrt{N}}\left[1,e^{-j\frac{2\pi}{\lambda}l\beta_{k}},...,e^{-j\frac{2\pi}{\lambda}l(N-1)\beta_{k}}\right], \stepcounter{equation}\tag{\theequation} \label{eq:arr_b}
\end{align*} 
where, $\alpha_{k}$ is the cosine of angle-of-arrival (AoA) and $\beta_{k}$ is the cosine of angle-of-departure (AoD). The term $\lambda$ is the carrier wavelength, whereas $l$ is the antenna separation. 

Maximizing P2 is equivalent to maximizing the received SNR at each robot position $q_{k}$~\eqref{eq:snr}. Assuming $\tilde{G}_{k}=\tilde{a}_{k}\tilde{b}_{k}^{T}$, and $\Phi_{k}=e^{\psi_{k}}\widehat{\Phi}_{k}$, this problem has a closed-form solution~\cite{IRSmmwave3}, which is given by:
\begin{align*} 
\psi_{k}^{*}=-\mbox{arg}\left(\left(\tilde{b}_{k}^{T}\right)^{H}\tilde{h}_{d,k}\right),\stepcounter{equation}\tag{\theequation} \label{eq:sol_alpha}
\end{align*} 
\begin{align*} 
\widehat{\Phi}_{k}^{*}=\mbox{diag}\left(e^{-j\mbox{arg}(g_{1,k})},...,e^{-j\mbox{arg}(g_{M,k})}\right),\stepcounter{equation}\tag{\theequation}\label{eq:sol_phi}
\end{align*} 
\begin{align*} 
w_{k}^{*}=\frac{\left(e^{\alpha_{k}^{*}}\sqrt{\rho d_{i,k}^{-\nu}}\tilde{h}_{r,k}^{H} \bar{\Phi}_{k}^{*} G_{k}+\sqrt{\rho d_{a,k}^{-\mu}}\tilde{h}_{d,k}^{H}\right)^{H}}{\lVert e^{\alpha_{k}^{*}}\sqrt{\rho d_{i,k}^{-\nu}}\tilde{h}_{r,k}^{H} \bar{\Phi}_{k}^{*} G_{k}+\sqrt{\rho d_{a,k}^{-\mu}}\tilde{h}_{d,k}^{H}\lVert_{2}},\stepcounter{equation}\tag{\theequation}\label{eq:sol_w}
\end{align*} 
where, $g_{k}=\sqrt{\rho d_{iak}^{-2}}\left(\tilde{h}_{r,k}^{*}\circ\tilde{a}_{k}\right)$ and $\left(\circ\right)$ denotes the elementwise product. By putting~\eqref{eq:sol_alpha}, \eqref{eq:sol_phi}, and~\eqref{eq:sol_w} into~\eqref{eq:snr}, we obtain the following optimal SNR expression for $q_{k}$:
\begin{align*} 
&\mbox{SNR}^{*}_{k}=\Big(N\lvert \rho \rvert \lvert \gamma \rvert^{2} \lVert \tilde{h}_{r,k}\lVert_{1}^{2}d_{i,k}^{-\nu}+\\
&2\sqrt{N} \lvert \rho \rvert \gamma \lVert \tilde{h}_{r,k}^{H}\lVert_{1} \lvert \tilde{b}_{k}^{T}\tilde{h}_{d,k} \rvert d_{i,k}^{-\nu/2}d_{a,k}^{-\mu/2}+\rho \lVert \tilde{h}_{d,k}\lVert_{2}^{2}d_{a,k}^{-\mu} \Big) \frac{p_{t}}{\sigma^{2}}=\\
&\Big(A d_{i,k}^{-\nu}+ B  d_{i,k}^{-\nu/2} d_{a,k}^{-\mu/2}+ Cd_{a,k}^{-\mu} \Big) \frac{p_{t}}{\sigma^{2}}.\stepcounter{equation}\tag{\theequation} \label{eq:snr_max}
\end{align*} 
In the last equality of~\eqref{eq:snr_max}, we have highlighted the dependence of $\mbox{SNR}^{*}_{k}$ on the robot position $q_{k}$ through the terms $d_{a,k}$ and $d_{i,k}$. However, $\nu$ and $\mu$ may rapidly change depending on the scattering environment and robot's position $q_{k}$. This can make the data rate model intractable for trajectory optimization in Section~\ref{sec:Traj_Opt}. For this reason, we proceed as follows: starting from~\eqref{eq:snr_max}, we assume $\nu$ and $\mu$ as constants and we estimate them, and other parameters, i.e., $A$, $B$, and $C$, from a set of measurements collected in a radio map. Then, in Section~\ref{sec:Traj_Opt}, we use the radio map information to address the dependence between the path loss exponents and the robot's position.

More precisely, given a set of channel measurements we compute~\eqref{eq:sol_alpha}, \eqref{eq:sol_phi}, and~\eqref{eq:sol_w} and construct a two-dimensional (2D) radio map that provides the averaged optimal $\mbox{SNR}^{*}$ for each position. Then, we can estimate $\widehat{A}\ge0$, $\widehat{B}\ge0$, $\widehat{C}\ge0$, $\widehat{\nu}\ge0$, and $\widehat{\mu}\ge0$ by fitting~\eqref{eq:snr_max} with the radio map. This procedure results in:
\begin{align*} 
\widehat{\mbox{SNR}}^{*}_{k}&=\Big(\widehat{A} d_{i,k}^{-\widehat{\nu}}+ \widehat{B}  d_{i,k}^{-\widehat{\nu}/2} d_{a,k}^{-\widehat{\mu}/2}+ \widehat{C}d_{a,k}^{-\widehat{\mu}} \Big) \frac{p_{t}}{\sigma^{2}},\stepcounter{equation}\tag{\theequation} \label{eq:snr_max_est}
\end{align*} 
where, $\widehat{A}$, $\widehat{B}$, $\widehat{C}$, $\widehat{\nu}$, and $\widehat{\mu}$ are the estimated parameters. This model has the advantages of analytical tractability of~\eqref{eq:snr_max} for trajectory optimization and capturing the dependence on the scattering environment. 

We can use~\eqref{eq:snr_max_est} in~\eqref{eq:avg_rate} to obtain an estimation of the maximum data rate resulting from the beamforming optimization:
\begin{align*} 
&\bar{\boldsymbol{\mbox{r}}}^{*}=\frac{1}{K}\sum_{k=0}^{K}\mbox{r}^{*}_{k}=\frac{\mbox{B}_{w}}{K}\sum_{k=0}^{K}\mbox{log}_{2}\left(1+\widehat{\mbox{SNR}}^{*}_{k}\right)=\\
&\frac{\mbox{B}_{w}}{K} \sum_{k=0}^{K}\mbox{log}_{2}\left(1+\Big(\widehat{A} d_{i,k}^{-\widehat{\nu}}+ \widehat{B}  d_{i,k}^{-\widehat{\nu}/2} d_{a,k}^{-\widehat{\mu}/2}+ \widehat{C} d_{a,k}^{-\widehat{\mu}} \Big) \frac{p_{t}}{\sigma^{2}}\right),\stepcounter{equation}\tag{\theequation} \label{eq:avg_rate2}
\end{align*} 
where, $\mbox{r}^{*}_{k}$ is the optimized data rate at position $q_{k}$. Finally, by replacing the LHS of~\eqref{con_Rate} with~\eqref{eq:avg_rate2}, we can decouple the beamforming and the trajectory optimization obtaining the following problem:
\begin{subequations}
\begin{align}
        P3:&\min_{\boldsymbol{q},\boldsymbol{\Phi},\boldsymbol{w}}E\label{opt3}\\
        \text{s.t.}& \;\bar{\boldsymbol{\mbox{r}}}^{*}\ge \mbox{r}_{min}\label{con_Rate3},\\
        		&~\eqref{Con_Dist},~\eqref{Con_Traj},~\eqref{Con_Ob},\nonumber
\end{align}
\end{subequations}
where, $E$ in~\eqref{opt3} is the robot energy consumption, which is given by~\eqref{eq:en_mot}.

\section{Trajectory Optimization}
\label{sec:Traj_Opt}
In this section, we provide an algorithm to solve problem $P3$ that, as introduced in Section~\ref{sec:Avg_Rate}, is a trajectory optimization problem. We first derive the following:
\begin{lemma}\label{lemma2}
Given $c_{1}\ge0$, $c_{2}\ge0$, and $c_{3}\ge0$, the objective function of $P3$~\eqref{opt3} is a convex function of $\boldsymbol{q}$.
\end{lemma}
\vspace{-0.5em}
\begin{proof}
We prove Lemma 1 by induction. As in~\eqref{eq:en_mot}, let $E|_{K=n}$ be the motion energy consumption of the robot when $K=n$: $E|_{K=n}=\sum_{k=1}^{n} c_{1} \frac{\lVert q_{k}-q_{k-1}\lVert_{2}^{2}}{\Delta_{t}}+c_{2}\lVert q_{k}-q_{k-1}\lVert_{2}+c_{3}\Delta_{t}$. We first prove that $E|_{K=1}$ is convex and then, by assuming that convexity holds for $E|_{K=n-1}$ we prove that $E|_{K=n}$ is a convex function of $\boldsymbol{q}=[q_{0},...,q_{n}]$. It is easy to show that $E|_{K=1}=c_{1} \frac{\lVert q_{1}-q_{0}\lVert_{2}^{2}}{\Delta_{t}}+c_{2}\lVert q_{1}-q_{0}\lVert_{2}+c_{3}\Delta_{t}$ is a convex function of $q_{0}$ and $q_{1}$ because it consists of the sum of two convex functions, i.e., $c_{1} \frac{\lVert q_{1}-q_{0}\lVert_{2}^{2}}{\Delta_{t}}$ and $c_{2}\lVert q_{1}-q_{0}\lVert_{2}$, and a constant term. Now, assume that $E|_{K=n-1}$ is convex, we consider $E|_{K=n}=E|_{K=n-1}+c_{1} \frac{\lVert q_{n}-q_{n-1}\lVert_{2}^{2}}{\Delta_{t}}+c_{2}\lVert q_{n}-q_{n-1}\lVert_{2}+c_{3}\Delta_{t}$. By following the same reasoning, we can observe that $E|_{K=n}$ is the sum of three convex functions of $\boldsymbol{q}=[q_{0},...,q_{n}]$: $E|_{K=n-1}$ that is convex by hypothesis, $\frac{\lVert q_{n}-q_{n-1}\lVert_{2}^{2}}{\Delta_{t}}$, and $c_{2}\lVert q_{n}-q_{n-1}\lVert_{2}$.
\end{proof}\vspace{-0.5em}
Thus, the objective function of $P3$ is a convex function of $\boldsymbol{q}$. However, $P3$ is non-convex because the LHS of~\eqref{con_Rate3} and~\eqref{Con_Ob} are not concave functions of $q_{k}$. For this reason, we perform a convex local approximation of these two constraints and solve the problem iteratively by using an SCO algorithm. Starting from constraint~\eqref{con_Rate3}, we have the following lemma:
\begin{lemma}\label{lemma1}
Given $\widehat{A}\ge0$, $\widehat{B}\ge0$, $ \widehat{C}\ge0$, $\widehat{\nu}\ge0$, and $\widehat{\mu}\ge0$, $\bar{\boldsymbol{\mbox{r}}}^{*}$ is a convex function of $d_{a,k}$ and $d_{i,k}$ with $k=0,...,K$.
\end{lemma}
\vspace{-0.5em}
\begin{proof}
See Appendix A.
\end{proof}
\vspace{-0.5em}

Thus, since any convex function can be lower-bounded by its first-order Taylor expansion, we have the following:
\begin{align*} 
\bar{\boldsymbol{\mbox{r}}}^{*}&\ge\bar{\boldsymbol{\mbox{r}}}_{apx}^{*}=\\
\frac{\mbox{B}_{w}}{K}&\sum_{k=0}^{K} \mbox{log}_{2}\left(1+\Big(\widehat{A} d_{i,0,k}^{-\widehat{\nu}}+ \widehat{B}  d_{i,0,k}^{-\widehat{\nu}/2} d_{a,0,k}^{-\widehat{\mu}/2}+\widehat{C} d_{a,0,k}^{-\widehat{\mu}} \Big) \frac{p_{t}}{\sigma^{2}}\right)+\\
&\nabla \bar{\boldsymbol{\mbox{r}}}^{*}|^{T}_{\big(d_{a,0,k},d_{i,0,k}\big)}
\begin{bmatrix}
d_{a,k}-d_{a,0,k}\\ 
d_{i,k}-d_{i,0,k}
\end{bmatrix},\stepcounter{equation}\tag{\theequation} \label{eq:avg_rate_appx}
\end{align*}

where, $\bar{\boldsymbol{\mbox{r}}}_{apx}^{*}$ is the first-order Taylor expansion of $\bar{\boldsymbol{\mbox{r}}}^{*}$ at expansion points $d_{a,0,k}$ and $d_{i,0,k}$. The term $\nabla \bar{\boldsymbol{\mbox{r}}}^{*}$ is the gradient of $\bar{\boldsymbol{\mbox{r}}}^{*}$ with respect to $d_{a,k}$ and $d_{i,k}$, which is given by:
\begin{align*} 
\nabla \bar{\boldsymbol{\mbox{r}}}^{*}=\frac{\mbox{B}_{w}}{K}\sum_{k=0}^{K}
\begin{bmatrix}
\frac{\Big(-\widehat{\nu}\widehat{A} d_{i,k}^{-\widehat{\nu}-1}-\widehat{\nu}/2\widehat{B}  d_{i,k}^{-\widehat{\nu}/2-1} d_{a,k}^{-\widehat{\mu}/2}\Big)\frac{p_{t}}{\sigma^{2}}}{\mbox{ln}2\Bigg(1+\Big(\widehat{A} d_{i,k}^{-\widehat{\nu}}+ \widehat{B}  d_{i,k}^{-\widehat{\nu}/2} d_{a,k}^{-\widehat{\mu}/2}+ \widehat{C} d_{a,k}^{-\widehat{\mu}} \Big)\frac{p_{t}}{\sigma^{2}}\Bigg)}\\ \\
\frac{\Big(-\widehat{\mu}\widehat{C} d_{a,k}^{-\widehat{\mu}-1}-\widehat{\mu}/2\widehat{B}  d_{i,k}^{-\widehat{\nu}/2} d_{a,k}^{-\widehat{\mu}/2-1}\Big)\frac{p_{t}}{\sigma^{2}}}{\mbox{ln}2\Bigg(1+\Big(\widehat{A} d_{i,k}^{-\widehat{\nu}}+ \widehat{B}  d_{i,k}^{-\widehat{\nu}/2} d_{a,k}^{-\widehat{\mu}/2}+ \widehat{C} d_{a,k}^{-\widehat{\mu}} \Big)\frac{p_{t}}{\sigma^{2}}\Bigg)}
\end{bmatrix}.\stepcounter{equation}\tag{\theequation} \label{eq:grad_rate}
\end{align*}
We now consider the following lemma:
\begin{lemma}\label{lemma3}
Given non-negative parameters $\widehat{A}$, $\widehat{B}$, $\widehat{C}$, $\widehat{\nu}$, and $\widehat{\mu}$, $\bar{\boldsymbol{\mbox{r}}}_{apx}^{*}$ is a concave function of $q_{k}$.
\end{lemma}
\vspace{-0.5em}
\begin{proof}
See Appendix B.
\end{proof}
\vspace{-0.5em}
Thus, in a small neighborhood of $d_{a,0,k}$ and $d_{i,0,k}$, we can derive $\bar{\boldsymbol{\mbox{r}}}_{apx}^{*}$ that is a concave function of $q_{k}$ and a lower bound of $\bar{\boldsymbol{\mbox{r}}}^{*}$. The same reasoning can be applied to constraints~\eqref{Con_Ob} leading to the following inequality:
\begin{align*} 
&\left(q_{k}-q_{c,o}\right)^{T}P_{o}^{-1}\left(q_{k}-q_{c,o}\right)\ge\\
&\left(q_{0,k}-q_{c,o}\right)^{T}P_{o}^{-1}\left(q_{0,k}-q_{c,o}\right)+\\
&\left(q_{0,k}-q_{c,o}\right)^{T}P_{o}^{-1}\left(q_{k}-q_{0,k}\right),\stepcounter{equation}\tag{\theequation} \label{eq:grad_obs}
\end{align*}
where, the LHS is the first-order Taylor expansion of~\eqref{Con_Ob} with respect to $q_{k}$ at local point $q_{0,k}$. This is an affine function of $q_{k}$.

Finally, by replacing~\eqref{con_Rate3} and~\eqref{Con_Ob} with~\eqref{eq:avg_rate_appx} and~\eqref{eq:grad_obs}, respectively, we can obtain a local convex approximation of $P3$ in a neighborhood of an initial feasible trajectory $\boldsymbol{q}_{0}$. As shown in Algorithm 1 (\textit{RMAP}), we can solve a sequence of local convex approximations of $P3$ that provides an upper bound to the solution of $P3$. Specifically, at each iteration $j$, \textit{RMAP} solves the following problem:
\begin{subequations}
\begin{align}
        P4: \min_{\boldsymbol{q}_{j}}&\sum_{k=1}^{K} c_{1} \frac{\lVert q_{j,k}-q_{j,k-1}\lVert_{2}^{2}}{\Delta_{t}}+c_{2}\lVert q_{j,k}-q_{j,k-1}\lVert_{2}+c_{3}\Delta_{t}\label{opt3}\\ 
        \text{s.t.}& \;\bar{\boldsymbol{\mbox{r}}}_{apx,j}^{*}\ge \mbox{r}_{min}\label{con_Rate4},\\
        		&\lVert q_{j,k}-q_{j,k-1} \lVert_{2}  \le D_{max}, \; \;\forall k, \label{Con_Dist4}\\
		&q_{0}=q_{s}, \; \; q_{k}=q_{d},\label{Con_Traj4}\\
        		&\lVert q_{j,k}-q_{j-1,k} \lVert_{2}  \le T_{k}, \; \;\forall k, \label{Con_Trust}\\
		&\left(q_{j-1,k}-q_{c,o}\right)^{T}P_{o}^{-1}\left(q_{j-1,k}-q_{c,o}\right)+\nonumber\\
&\left(q_{j-1,k}-q_{c,o}\right)^{T}P_{o}^{-1}\left(q_{j,k}-q_{j-1,k}\right)\ge 1,  \; \forall k,o, \label{Con_Ob4}
\end{align}
\end{subequations}
where, $\boldsymbol{q}_{j}=[q_{j,0},q_{j,1},...,q_{j,K}]$ and $\boldsymbol{q}_{j-1}=[q_{j-1,0},q_{j-1,1},...,q_{j-1,K}]$ are the solutions of $P4$ at iteration $j$ and $j-1$, respectively. More precisely, $\boldsymbol{q}_{j-1}$ represents the local point at which the approximations at iteration $j$ of constraints~\eqref{con_Rate3} and~\eqref{Con_Ob} are computed. These approximations are valid in a trust region of $\boldsymbol{q}_{j-1}$ that is defined by constraint~\eqref{Con_Trust}. The trust region size $T_{k}$ may differ by position $q_{k}$. Note that the expansion points of $\bar{\boldsymbol{\mbox{r}}}^{*}$ can be obtained from $q_{j-1,k}$, as $d_{a,j-1,k}=\sqrt{(z_{r}-z{a})^{2}+\lVert q_{j-1,k}-q_{a}\lVert^{2}_{2}}$ and $d_{i,j-1,k}=\sqrt{(z_{r}-z{i})^{2}+\lVert  q_{j-1,k}-q_{i}\lVert^{2}_{2}}$. Problem $P4$ is convex and it can be solved quickly by interior-point methods.

As introduced in Section~\ref{sec:Avg_Rate}, $\bar{\boldsymbol{\mbox{r}}}^{*}$ and its approximation ($\bar{\boldsymbol{\mbox{r}}}_{apx}^{*}$) can still diverge from the true data rate, especially when abrupt LOS-NLOS transitions occur. Thus, in \textit{RMAP} we introduce a solution update mechanism that differs from conventional SCO-based algorithms. The goal is to keep the feasibility of the solution at iteration $j$ also with respect to the measured data rate obtained from the radio map and not only to its convex approximation ($\bar{\boldsymbol{\mbox{r}}}_{apx,j}^{*}$). More precisely, let $\boldsymbol{\mbox{r}}_{map,j}$ be a vector of $K$ elements, each of which consists of the measured data rate with optimized beamforming vectors at $q_{j,k}$. Let $\bar{\boldsymbol{\mbox{r}}}_{map,j}$ be the average of $\boldsymbol{\mbox{r}}_{map,j}$ along the trajectory. Then, at iteration $j$, \textit{RMAP} updates the solution, only if $\bar{\boldsymbol{\mbox{r}}}_{map,j}\ge r_{min}$, otherwise it keeps the previous solution, i.e., $\boldsymbol{q}_{j}=\boldsymbol{q}_{j-1}$. Furthermore, to obtain following feasible solutions, the algorithm reduces the trust-region $T_{k}$ of a factor $0\le \tau<1$, where $k$ represents the position at which the measured data rate from the radio map drops the most with respect to the previous trajectory solution: $k=\argmax\limits_k\mbox{r}_{map,j-1,k}-\mbox{r}_{map,j,k}$. 
\begin{algorithm}
\algsetup{linenosize=\tiny}
\caption{\textit{Radio Map Assisted Planning (RMAP)}}
\label{alg:Algo}
\begin{algorithmic}[1]
  \scriptsize
  \item[\textbf{\textit{Initial solution}}:]
  \STATE{j=0}
  \STATE{Find an initial feasible solution $\boldsymbol{q}_{j}$}
  \STATE{Compute the motion energy consumption $E_{j}$ corresponding to $\boldsymbol{q}_{j}$ as in~\eqref{opt3}}  
  \item[\textbf{\textit{SCO}}:]
  \REPEAT {
  	\STATE{$j=j+1$}  
 	 \STATE{Obtain $\boldsymbol{q}_{j}$ and $E_{j}$ by solving $P4$ with local points $\boldsymbol{q}_{j-1}$}	 
	 \IF{$\bar{\boldsymbol{\mbox{r}}}_{map,j} < r_{min}$} 
		\STATE{$\boldsymbol{q}_{j}=\boldsymbol{q}_{j-1}$ and $E_{j}=E_{j-1}$}
		\STATE{$T_{k}=\tau T_{k}$ with, $0\le \tau <1$ and $k=\argmax\limits_k\mbox{r}_{map,j-1,k}-\mbox{r}_{map,j,k}$}
	 \ELSE
		\IF{$\frac{E_{j}-E_{j-1}}{E_{j-1}}\le \epsilon$} 
			\BREAK
		\ENDIF 
	 \ENDIF 
	 } \UNTIL{$j\ge N_{it}$}
\end{algorithmic}
\end{algorithm}

Hence, the algorithm maintains the feasibility with respect to both $\bar{\boldsymbol{\mbox{r}}}^{*}$ and $\bar{\boldsymbol{\mbox{r}}}_{map}$. The analytical tractability of the former is used to optimize the trajectory and understand the behavior of the data rate with respect to the distances between the robot, the AP, and the IRS. The latter ($\bar{\boldsymbol{\mbox{r}}}_{map}$) is used to capture the NLOS and LOS transitions created by the obstacles. The algorithm stops if the sequence of solutions converges or when a maximum number of iterations ($N_{it}$) is reached. More precisely, we can prove that the algorithm converges and, under some conditions, it converges to a Karush-Kuhn-Tucker (KKT) point of $P3$.
\vspace{-0.5em}
\begin{theorem}\label{theo2}
\textit{RMAP} provides a non-increasing and convergent sequence of solutions. Moreover, if at each iteration $j$ we have that $\bar{\boldsymbol{\mbox{r}}}_{map,j}\ge r_{min}$, \textit{RMAP} converges to a KKT point of $P3$.
\end{theorem}
\vspace{-0.5em}
\begin{proof}
The sequence of solutions provided by \textit{RMAP} is non increasing because, the solution of $P4$ at iteration $j-1$, $\boldsymbol{q}_{j-1}$, is a feasible solution of minimization problem $P4$ at iteration $j$. For the rest of the proof, see Appendix~C and Appendix~D.
\end{proof}
\vspace{-0.5em}
\begin{prop}\label{prop1}
If the average measured data rate corresponding to the initial solution satisfies $\bar{\boldsymbol{\mbox{r}}}_{map,0}\ge r_{min}$, then the solution to which \textit{RMAP} converges satisfies this constraint as well.
\end{prop}
\vspace{-0.5em}
This is a direct consequence of Lines 7-9 of \textit{RMAP} and Theorem 1. Namely, if $\bar{\boldsymbol{\mbox{r}}}_{map,j} < r_{min}$ then $\boldsymbol{q}_{j}=\boldsymbol{q}_{j-1}$. Moreover, assuming that the initial solution satisfies $\bar{\boldsymbol{\mbox{r}}}_{map,0} \ge r_{min}$ and \textit{RMAP} converges to a solution, this solution must satisfy the above constraint.

In general, the quality of a solution of SCO-based algorithms also depends on the initial solution. In this work, we obtain $\boldsymbol{q}_{0}$ by using a graph-based method. Specifically, we compute the shortest path on a time-expanded graph as done in~\cite{MAPP}. The edges and vertices of the graph are defined on a discrete set of positions that are free from obstacles. In each timeslot, a robot may either stay at a vertex or move to an adjacent one. The distance between vertices is set according to the robot's maximum speed. On this graph, the costs of the edges are set to generate two different initial solutions. The first one minimizes the motion energy consumption (ME), whereas the second solution maximizes the data rate (MR). The radio map is used to obtain the SNR and the data rate for the positions corresponding to vertices and edges. Then, \textit{RMAP} uses the minimum energy initial solution if it is feasible and the maximum data rate initial solution, otherwise. If the latter is not feasible, the algorithm declares infeasibility.

\begin{figure}[tb]
	\centering
	\includegraphics[width=8.5cm]{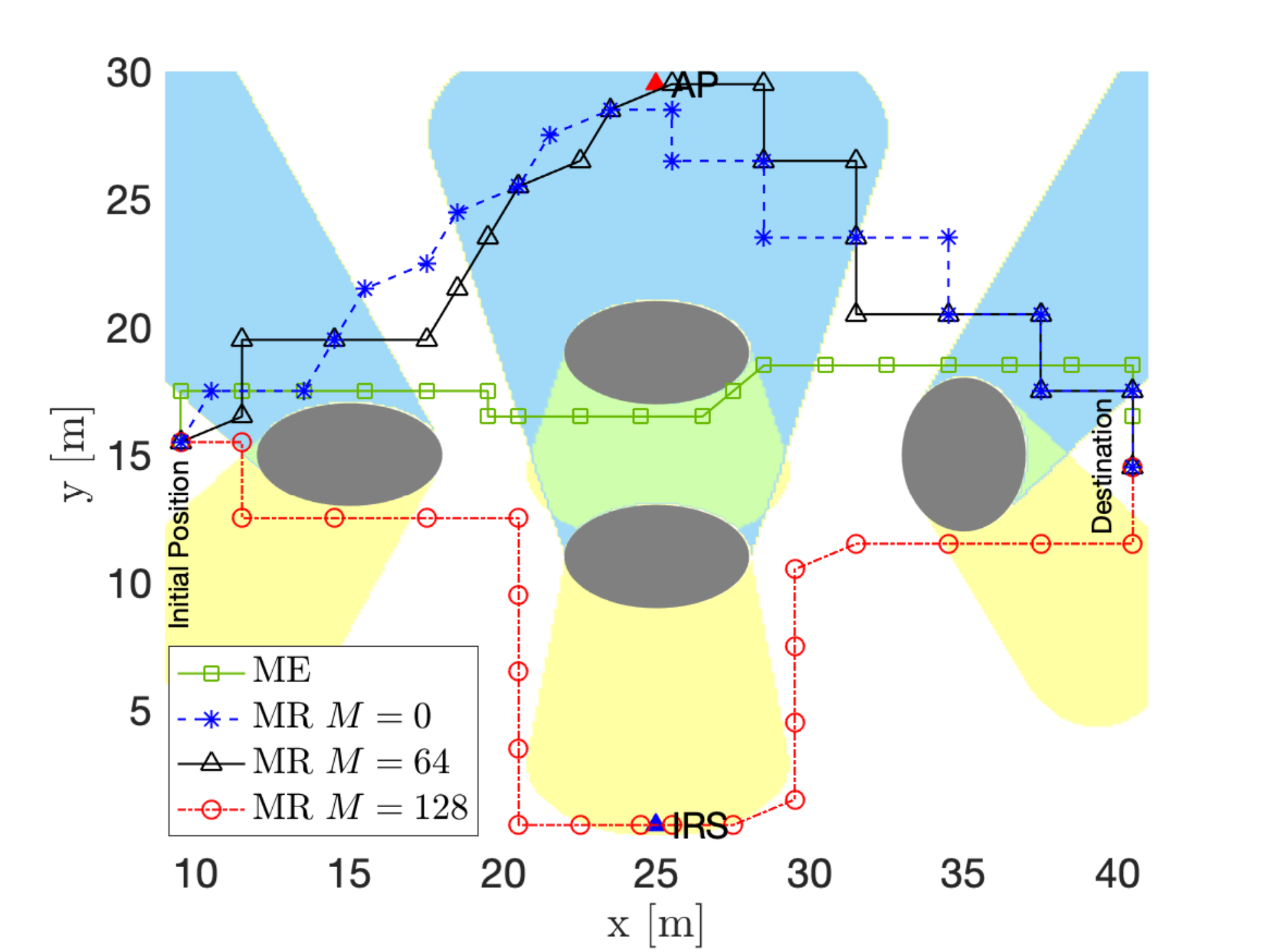}
	\caption[]{Minimum energy (ME) and maximum data rate (MR) initial solutions for $K=30$ and several values of $M$. Yellow, blue, and green shaded positions are in NLOS with respect to the AP, the IRS, and both, respectively. The positions in the white area are in LOS with respect to both the IRS and the AP.}
	\label{fig:Init}
\end{figure}

\section{Numerical Results}
\label{sec:Res}
In this section, we provide a numerical validation of \textit{RMAP} for solving $P3$. For our simulations, we consider a $50 \times 30$\,m$^{2}$ rectangular-shaped indoor scenario. The robot's starting position is $[9.5,15.5]$, whereas the destination is $[40.5,14.5]$. There are an AP and an IRS placed at [25,30] and [25,0], respectively, operating in the $60$~GHz band as in~\cite{mmInd2}, with bandwidth $B_{w}=200$ MHz. The height of the AP is $5$\,m, whereas we set the heights of the IRS and the robot's antenna to $2.5$\,m and $0.5$\,m, respectively. For sake of clarity, we first present results for a scenario consisting of four ellipse obstacles that are placed as in Fig.~\ref{fig:Init}, represented by grey shaded areas (\textit{base scenario}). This scenario includes several robot-AP and robot-IRS channel conditions, i.e., LOS and NLOS positions. Then, we present results that are averaged over ten instances in which $20$ obstacles are randomly placed (\textit{random scenario}). In both scenarios, the length, width, and height of obstacles are $6$\,m, $4$\,m, and $2$\,m, respectively. 
\begin{figure}[!tbp]
  \begin{subfigure}[b]{0.45\textwidth}
    \includegraphics[width=8.5cm]{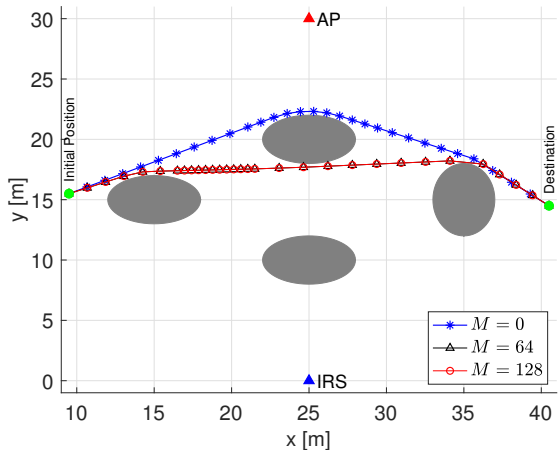}
    \caption{$r_{min}=2.0$\; Gbps.}
    \label{fig:path2}
  \end{subfigure}
  \hfill
  \begin{subfigure}[b]{0.45\textwidth}
    \includegraphics[width=8.5cm]{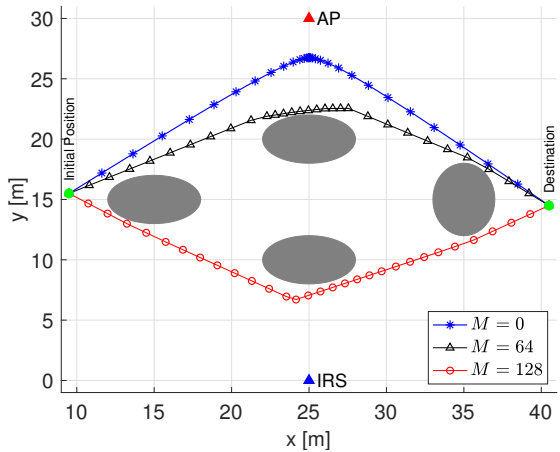}
    \caption{$r_{min}=2.5$\; Gbps.}
    \label{fig:path2_5}
  \end{subfigure}
  \caption{Robot trajectories to which \textit{RMAP} converges for $K=30$, and several values of $M$ and $r_{min}$.}
\end{figure}
\begin{figure}[!tbp]
  \begin{subfigure}[b]{0.45\textwidth}
    \includegraphics[width=8.5cm]{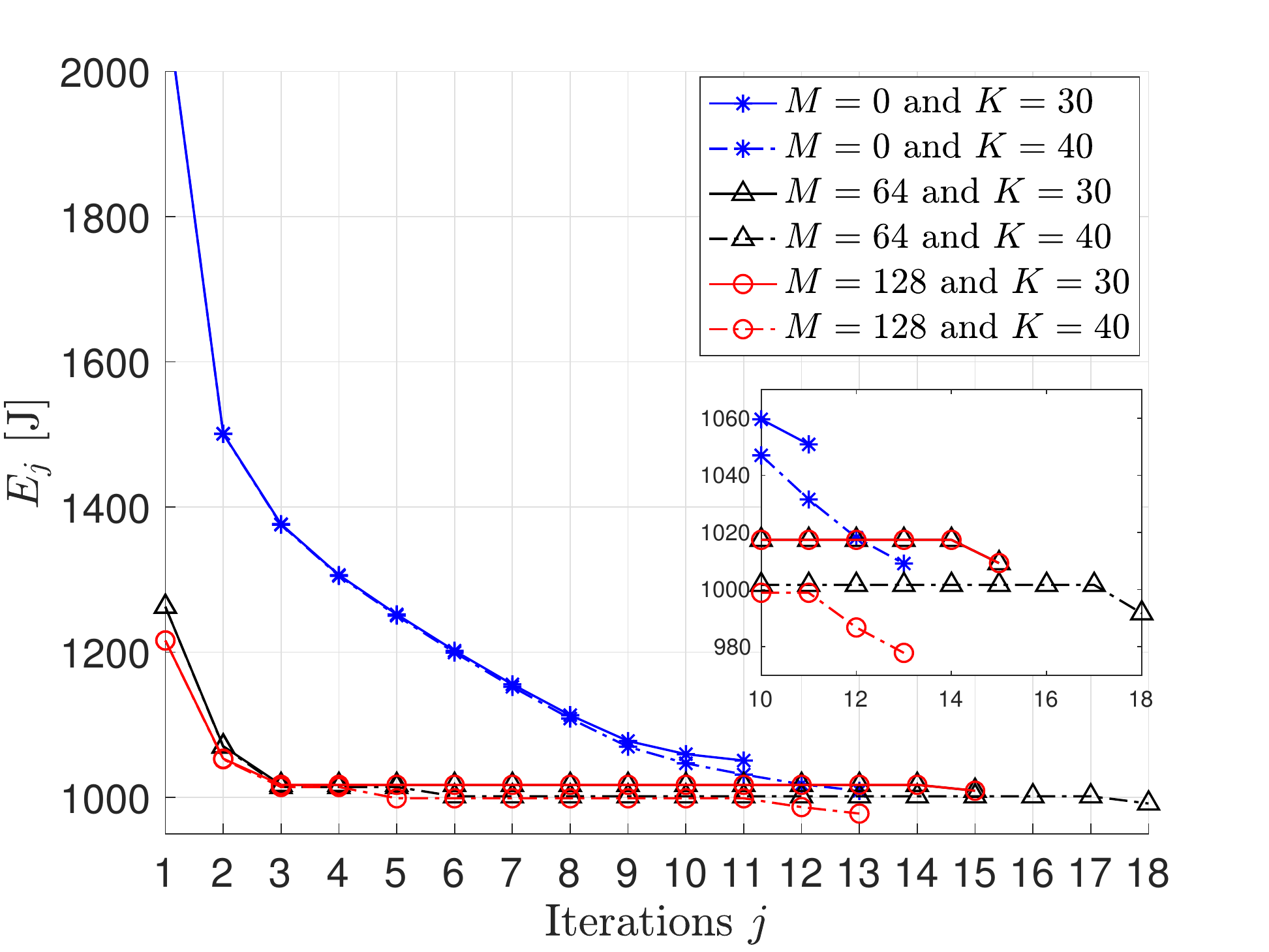}
    \caption{$r_{min}=2.0$\; Gbps.}
    \label{fig:it2}
  \end{subfigure}
  \hfill
  \begin{subfigure}[b]{0.45\textwidth}
    \includegraphics[width=8.5cm]{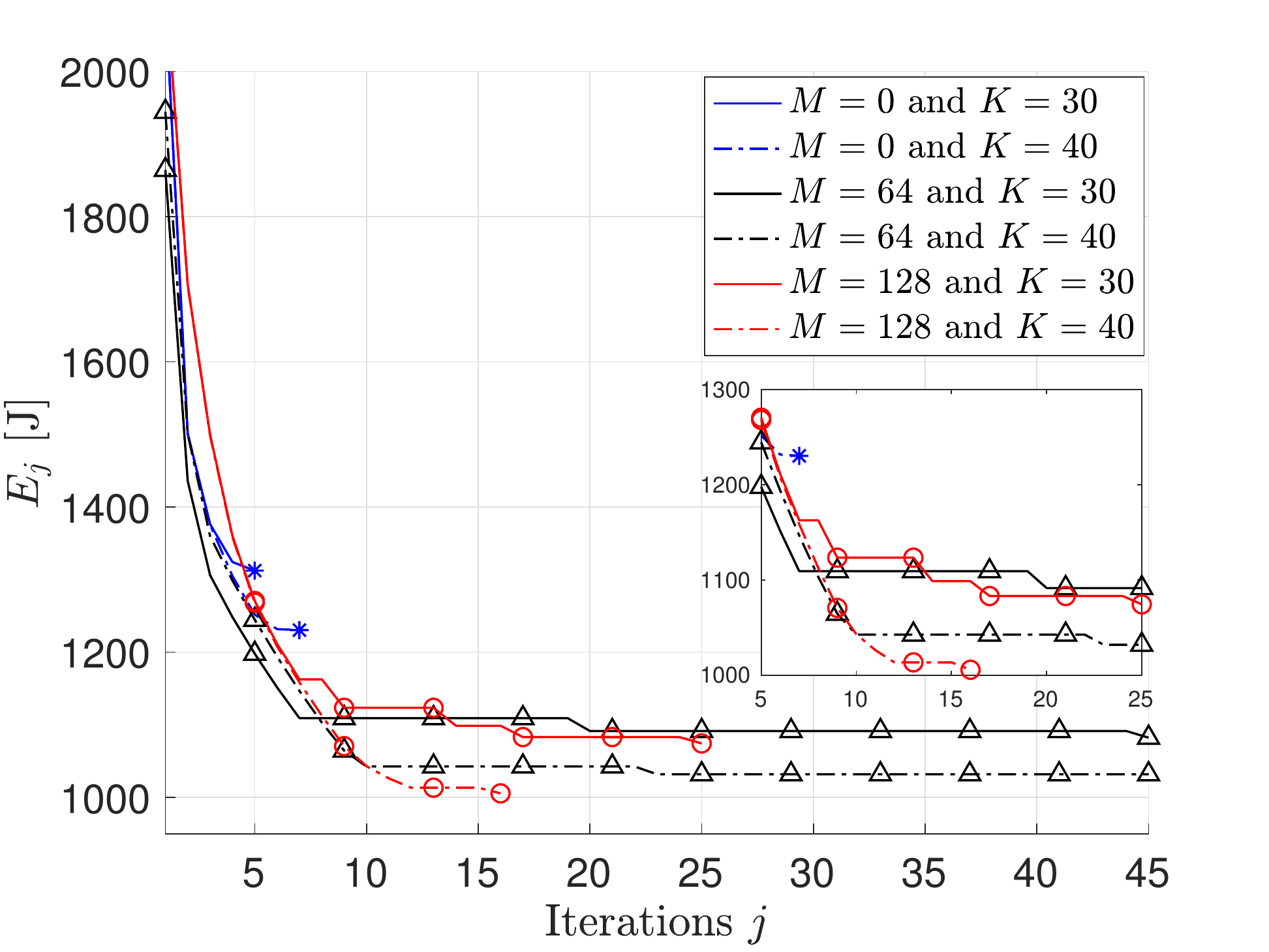}
    \caption{$r_{min}=2.5$\; Gbps.}
    \label{fig:it2_5}
  \end{subfigure}
  \caption{Energy consumption corresponding to the sequence of solutions $\boldsymbol{q}_{j}$ provided by \textit{RMAP} for several values of $M$, $r_{min}$, and $K$.}
\end{figure}

Similar to~\cite{mmInd2}, the path loss at a reference distance of $1$\,m is $68$ dB, and the path loss exponent of the robot-AP and robot-IRS channels are set to $2$ for LOS, and $4.5$ for NLOS. Without loss of generality the antenna gain of the reflective elements are set to $0$\,dBi. Moreover, we set the transmit and the noise powers to $20$\,dBm, and $-80$\,dBm, respectively. We show the results, for several values of $M$ and $r_{min}$, and, unless otherwise specified, we set the following parameters values: $K=30$, $\tau=0.5$, $N=16$, $\Delta_{t}=1$\,s, $v_{max}=3$\,m/s, $d_{s}=1.35$, $N_{it}=100$, $\epsilon=0.01$, $T=1$\,m, $c_{1}=4.39$, $c_{2}=24.67$, and $c_{3}=14.77$~\cite{CoCP}. Finally, to derive~\eqref{eq:snr_max_est}, we estimate $\widehat{A}\ge0$, $\widehat{B}\ge0$, $\widehat{C}\ge0$, $\widehat{\nu}\ge0$, and $\widehat{\mu}\ge0$. These parameters are obtained by fitting~\eqref{eq:snr_max} on a radio map by solving a non-linear least squares problem. The radio map is obtained from the average of $10.000$ channel measurements on a grid of $500\times 300$ points.

\subsection{Base Scenario}
\label{sec:fixed}
In Fig.~\ref{fig:Init}, we first show ME and MR initial solutions for several values of $M$. Note that the MR initial solution considers the trajectory that maximizes the data rate. Such a trajectory tends to avoid NLOS areas with respect to either the AP or to the IRS depending on the number of reflective elements ($M$) of the latter. 

In Fig.~\ref{fig:path2} we show robot trajectories resulting from \textit{RMAP} for $K=30$, $r_{min}=2.0$\,Gbps, and several values of $M$. For $M=0$, we can observe that the robot avoids NLOS areas with respect to the AP. Specifically, \textit{RMAP} uses initial solution MR. When $M$ increases, the IRS enhances the coverage such that the robot can find a trajectory with lower energy consumption ($E$) by using initial solution ME. The resulting trajectory crosses the NLOS area with respect to both the AP and the IRS. Note that, for $r_{min}=2.0$\,Gbps, values of $M$ that are higher than $64$ do not provide further gain, thus the trajectories for $M=64$ and $M=128$ coincide. This is not true when $r_{min}=2.5$\,Gbps, for which the trajectories are shown in Fig.~\ref{fig:path2_5}. More precisely, when $r_{min}$ increases, \textit{RMAP} selects the MR initial solutions for all the values of $M$ and the resulting paths are either closer to the AP or the IRS to improve the coverage and increase the data rate. For $M=0$ and $M=64$ the robot trajectories avoid completely NLOS areas with respect to the AP, whereas, for $M=128$, the paths that are closer to the IRS provide higher data rates. We can also observe that the robot decreases the speed when the data rate is higher. Specifically, in LOS positions that are closer to the AP and the IRS, the robot travels for a smaller distance in each timeslot to exploit better coverage.

In general, $E$ increases for higher values of $r_{min}$ and decreases when $M$ and $K$ increase. This is more prominent in Fig.~\ref{fig:it2} and Fig.~\ref{fig:it2_5} where we show energy consumption $E_{j}$ corresponding to the sequence of solutions $\boldsymbol{q}_{j}$ provided by \textit{RMAP} for $r_{min}=2.0$\,Gbps and $r_{min}=2.5$\,Gbps, respectively. First, we can observe that, by increasing $M$, \textit{RMAP} converges to paths with lower $E$. The energy consumption also decreases when $K$ increases. Specifically, for a fixed value of $\Delta_{t}$, higher values of $K$ correspond to longer deadlines and the robot can decrease the speed to reach the destination. Then, as described by~\eqref{eq:en_mot}, lower speeds correspond to lower values of $E$. Moreover, we can observe that $E_{j}$ is non-increasing and \textit{RMAP} converges in few iterations. However, as explained and shown better in the following section, the number of iterations within which \textit{RMAP} converges depends on the value of $\tau$. 

\begin{figure}[!tbp]
  \centering
  \begin{minipage}[b]{0.45\textwidth}
    \includegraphics[width=8.5cm]{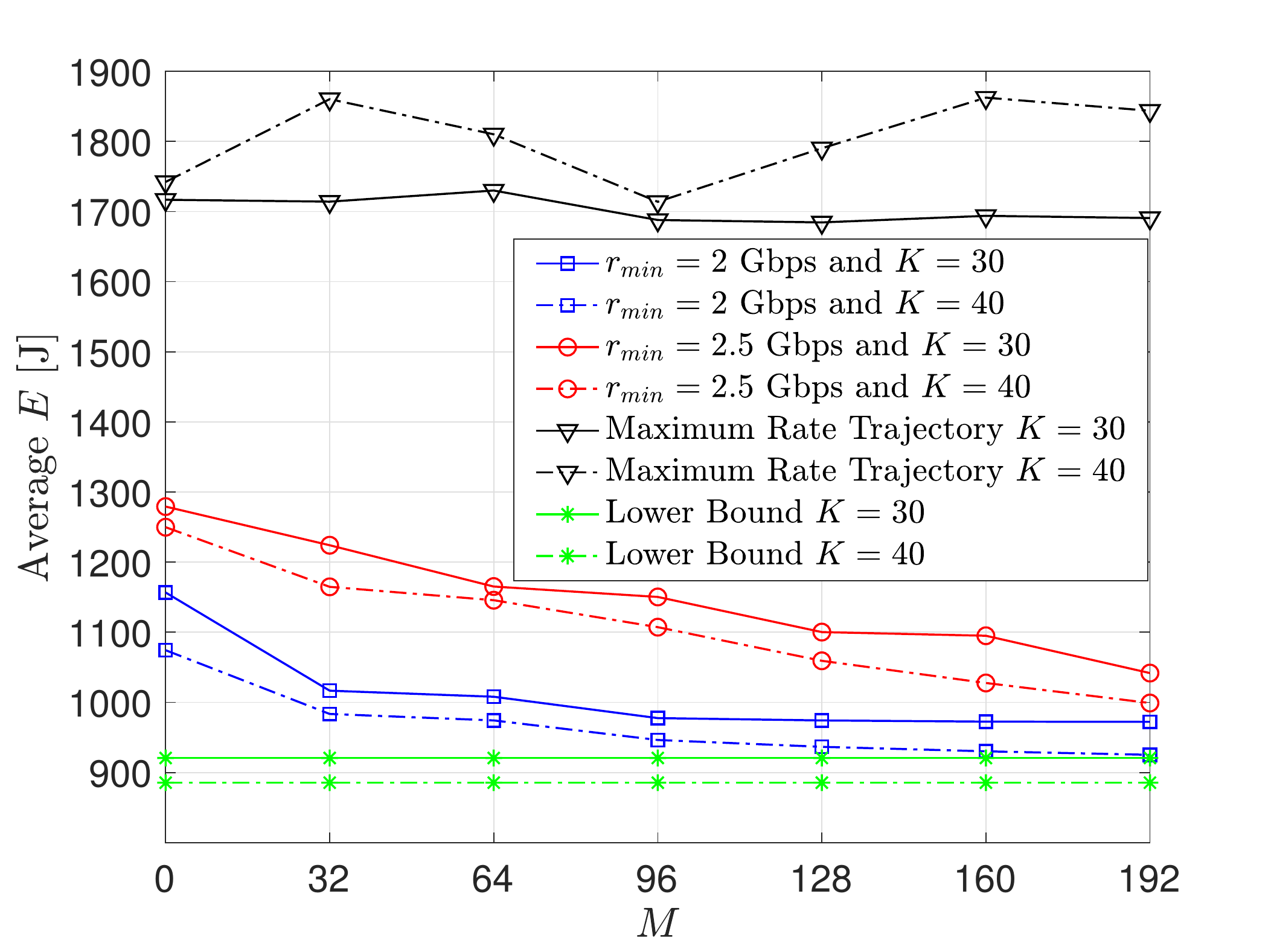}
    \caption{Average $E$ corresponding to solutions of \textit{RMAP} and maximum data rate trajectories for several values of $M$, $r_{min}$, and $K$. Moreover, we show lower bounds to average $E$ for various values of $K$.}
    \label{fig:En_M}
  \end{minipage}
  \hfill
  \begin{minipage}[b]{0.45\textwidth}
    \includegraphics[width=8.5cm]{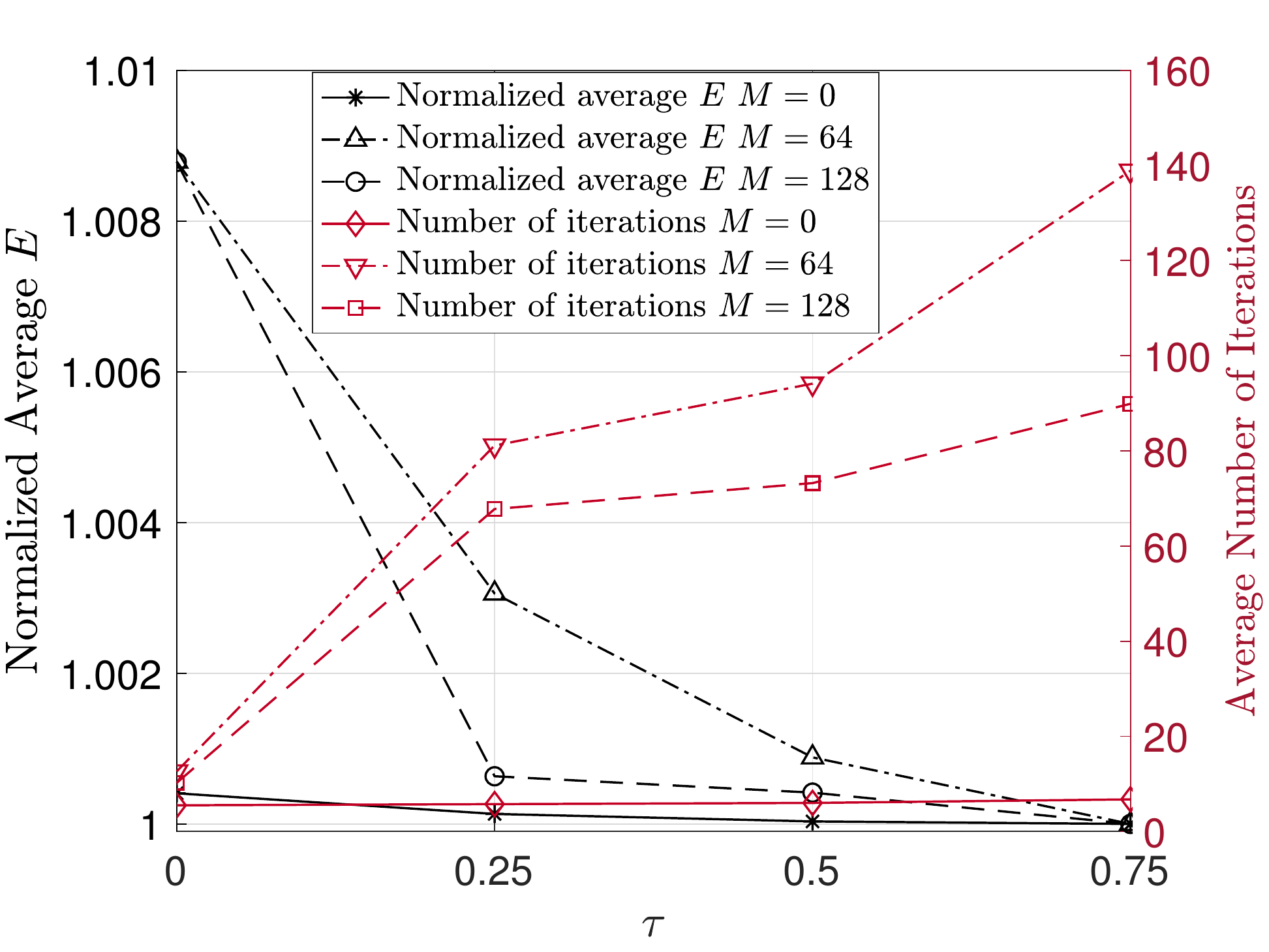}
    \caption{Normalized average $E$ (black) and number of iterations (red) within which \textit{RMAP} converges for several values of $M$, $\tau$, $K=30$, and $r_{min}=2.5$\,Gbps.}
    \label{fig:ratio}
  \end{minipage}
\end{figure}
\subsection{Random Scenario}
\label{sec:ran}
In this section, we present results that are averaged over $10$ instances in which $N_{o}=20$ obstacles are randomly placed. The area, the obstacles dimension, and the robot's starting position and destination are the same that are used for Section~\ref{sec:fixed}. In Fig.~\ref{fig:En_M}, we show the average energy consumption corresponding to the trajectory resulting from \textit{RMAP} for several values of $M$, $K$, and $r_{min}$. Moreover, we show the average $E$ corresponding to robot trajectories that maximize the data rate under time and collision avoidance constraints. In addition, Fig.~\ref{fig:En_M} depicts lower bounds to the average energy consumption that are computed by solving a relaxed version of problem $P3$, where obstacle avoidance~\eqref{Con_Ob} and QoS~\eqref{con_Rate3} constraints are relaxed. This relaxation results in a convex optimization problem that provides lower bounds to solutions of $P3$. Conversely, it is important to highlight that \textit{RMAP} provides upper bounds to $P3$.

Maximum data rate trajectories are obtained by solving a modified version of P3, where the LHS of~\eqref{con_Rate3} represents the objective function of the problem. This problem is solved by using an SCO-based algorithm where, as done in Section~\ref{sec:Traj_Opt} for $P3$, the objective function and the LHS of~\eqref{Con_Ob} are approximated by convex functions. Note that maximum-rate trajectories are feasible solutions of $P3$, whereas trajectories resulting from computing the lower bounds may be not.

In Fig.~\ref{fig:En_M}, we can observe that, for both $r_{min}=2.0$\,Gbps and $r_{min}=2.5$\,Gbps, \textit{RMAP} can reduce dramatically the average $E$ with respect to the maximum data rate approach. The gain is close to $100$\% for $r_{min}=2.0$\,Gbps and $K=40$. Moreover, Fig.~\ref{fig:En_M} shows that the average $E$ resulted from \textit{RMAP} decreases and approaches the lower bound as $M$ increases. Specifically, by increasing the number of reflective elements at the IRS we enhance the coverage, and the robot can find a higher number of feasible trajectories. However, while for $r_{min}=2.5$\,Gbps, increasing $M$ results in a monotonic decrease of $E$, for $r_{min}=2.0$\,Gbps, we note that above a certain threshold ($M \ge 32$) increasing the value of $M$ does not provide significant gains. Additional gain can be obtained by increasing the values of $K$ as also explained in Section~\ref{sec:fixed}. These observations do not hold for maximum data rate trajectories for which the solutions have not monotonic decreasing behaviors with respect to $M$.

Finally, in Fig.~\ref{fig:ratio} we show the effects of parameter $\tau$ on the number of iterations within which \textit{RMAP} converges. Moreover, we show the average $E$ corresponding to the solutions within which \textit{RMAP} converges for several values of $\tau$. These are normalized to the solutions that are obtained for $\tau=0.75$. As explained in Section~\ref{sec:Traj_Opt}, in each iteration of \textit{RMAP} for which the solution does not satisfy $\bar{\boldsymbol{\mbox{r}}}_{map,j}\ge r_{min}$, we multiply the trust region size (of the position where the data rate drops the most) by $\tau$. Specifically, when $\tau$ is smaller, the sizes of the trust regions may decrease faster leading to a faster algorithm convergence. This can be observed in Fig.~\ref{fig:ratio}, which shows the tradeoff between the number of iterations that \textit{RMAP} needs to converge and the quality of the solution. The latter improves when $\tau$ grows, which leads to lower values of average $E$. However, when $\tau=0$, \textit{RMAP} provides solutions in fewer iterations and negligible loss ($\le1\%$) with respect to the ones provided for $\tau=0.75$.


\section{Conclusion}
\label{sec:Conc}
In this work, we have proposed a novel robot trajectory optimization problem with QoS constrained communications for minimizing the motion energy consumption. The robot must avoid collisions with obstacles, reach the destination within a deadline, and transmit data to an AP operating at mm-wave frequency bands and assisted by an IRS. The uplink transmission is subject to a minimum average data rate. To the best of our knowledge, energy-efficient trajectory planning problems for wirelessly connected robots have not been considered for mm-wave communications. We have proposed a solution that accounts for the challenging signal propagation conditions at such high frequencies and the mutual dependence between the channel conditions and the robot trajectory. Specifically, we have decoupled the beamforming and the trajectory optimization problems by exploiting the mm-wave propagation characteristics. The latter is solved by an SCO-based algorithm (\textit{RMAP}) for which the convergence is proved. \textit{RMAP} can deal with sudden data rate drops due to LOS-NLOS transitions by using the information that is stored in a radio map. Given this information, \textit{RMAP} can find trajectories that avoid obstacles and poorly connected areas for satisfying data rate requirements. 

We have shown trajectories and corresponding energy consumptions at which the algorithm converges for several scenarios and system parameters. The algorithm converges in few iterations to solutions that approach the lower bound to the energy consumption, which is dramatically reduced with respect to trajectories that maximize the data rate. Finally, we have shown that, by increasing the number of IRS's reflective elements, we can improve the coverage and reduce the energy consumption of wirelessly connected robots. Thus, given the negligible power consumption of passive IRSs, they represent powerful solutions to enhance the energy efficiency of fully connected and autonomous factories.

\appendices
\section{}
\label{sec:ApA}
To prove Lemma 2, we prove that $\mbox{r}_{k}^{*}$, which is the estimated data rate at position $q_{k}$, is a convex function of $d_{i,k}$ and $d_{a,k}$. Then, $\bar{\boldsymbol{\mbox{r}}}^{*}$ is convex because it is a sum of convex functions. We first compute the partial derivatives of $\mbox{r}_{k}^{*}$ with respect to $d_{i,k}$ and $d_{a,k}$. These are given by:
\begin{align*} 
\frac{\partial\mbox{r}_{k}^{*}}{\partial d_{i,k}}=\frac{\Big(-\widehat{\nu}\widehat{A} d_{i,k}^{-\widehat{\nu}-1}-\widehat{\nu}/2\widehat{B}  d_{i,k}^{-\widehat{\nu}/2-1} d_{a,k}^{-\widehat{\mu}/2}\Big)\frac{p_{t}}{\sigma^{2}}} {F_{k}},\stepcounter{equation}\tag{\theequation} \label{eq:rate_dev1}
\end{align*}
\begin{align*} 
\frac{\partial \mbox{r}_{k}^{*}}{\partial d_{a,k}}=\frac{\Big(-\widehat{\mu}\widehat{C} d_{a,k}^{-\widehat{\mu}-1}-\widehat{\mu}/2\widehat{B}  d_{i,k}^{-\widehat{\nu}/2} d_{a,k}^{-\widehat{\mu}/2-1}\Big)\frac{p_{t}}{\sigma^{2}}}{F_{k}},\stepcounter{equation}\tag{\theequation} \label{eq:rate_dev2}
\end{align*}
where, $F_{k}=\mbox{ln}(2)\Bigg(1+\Big(\widehat{A} d_{i,k}^{-\widehat{\nu}}+ \widehat{B}  d_{i,k}^{-\widehat{\nu}/2} d_{a,k}^{-\widehat{\mu}/2}+ \widehat{C} d_{a,k}^{-\widehat{\mu}} \Big)\frac{p_{t}}{\sigma^{2}}\Bigg)>0$. Then, the second order partial derivatives are given by:
\begin{align*} 
\frac{\partial^{2} \mbox{r}_{k}^{*}}{\partial d_{i,k}^{2}}&=\frac{\Big(\widehat{\nu}(\widehat{\nu}+1)\widehat{A} d_{i,k}^{-\widehat{\nu}-2}+\widehat{\nu}/2(\widehat{\nu}/2+1)\widehat{B}  d_{i,k}^{-\widehat{\nu}/2-2} d_{a,k}^{-\widehat{\mu}/2}\Big)F_{k}\frac{p_{t}}{\sigma^{2}}}{F_{k}^{2}}\\
&-\frac{\mbox{ln}(2)\Big(-\widehat{\nu}\widehat{A} d_{i,k}^{-\widehat{\nu}-1}-\widehat{\nu}/2\widehat{B}  d_{i,k}^{-\widehat{\nu}/2-1} d_{a,k}^{-\widehat{\mu}/2}\Big)^{2}\frac{p_{t}^{2}}{\sigma^{4}}}{F_{k}^{2}},\stepcounter{equation}\tag{\theequation} \label{eq:rate_dev11}
\end{align*}
\begin{align*} 
\frac{\partial^{2} \mbox{r}_{k}^{*}}{\partial d_{a,k}^{2}}&=\frac{\Big(\widehat{\mu}(\widehat{\mu}+1)\widehat{C} d_{a,k}^{-\widehat{\mu}-2}+\widehat{\mu}/2(\widehat{\mu}/2+1)\widehat{B}  d_{i,k}^{-\widehat{\nu}/2} d_{a,k}^{-\widehat{\mu}/2-2}\Big)F_{k}\frac{p_{t}}{\sigma^{2}}}{F_{k}^{2}}\\
&-\frac{\mbox{ln}(2)\Big(-\widehat{\mu}\widehat{C} d_{a,k}^{-\widehat{\mu}-1}-\widehat{\mu}/2\widehat{B}  d_{i,k}^{-\widehat{\nu}/2} d_{a,k}^{-\widehat{\mu}/2-1}\Big)^{2}\frac{p_{t}^{2}}{\sigma^{4}}}{F_{k}^{2}},\stepcounter{equation}\tag{\theequation} \label{eq:rate_dev22}
\end{align*}
\begin{align*} 
\frac{\partial^{2} \mbox{r}_{k}^{*}}{\partial d_{i,k} \partial d_{a,k}}&=\frac{\Big((\widehat{\mu}/2)(\widehat{\nu}/2)\widehat{B}  d_{i,k}^{-\widehat{\nu}/2-1} d_{a,k}^{-\widehat{\mu}/2-1}\Big)F_{k}\frac{p_{t}}{\sigma^{2}}}{F_{k}^{2}}-\\
&\frac{\mbox{ln}(2)\Big(-\widehat{\nu}\widehat{A} d_{i,k}^{-\widehat{\nu}-1}-\widehat{\nu}/2\widehat{B}  d_{i,k}^{-\widehat{\nu}/2-1} d_{a,k}^{-\widehat{\mu}/2}\Big)\frac{p_{t}^{2}}{\sigma^{4}}}{F_{k}^{2}} \times\\
&\frac{\Big(-\widehat{\mu}\widehat{C} d_{a,k}^{-\widehat{\mu}-1}-\widehat{\mu}/2\widehat{B}  d_{i,k}^{-\widehat{\nu}/2} d_{a,k}^{-\widehat{\mu}/2-1}\Big)\frac{p_{t}^{2}}{\sigma^{4}}}{F_{k}^{2}}.\stepcounter{equation}\tag{\theequation} \label{eq:rate_dev12}
\end{align*}
We observe that $\frac{\partial^{2} \mbox{r}_{k}^{*}}{\partial d_{i,k}^{2}}>0$, $\frac{\partial^{2} \mbox{r}_{k}^{*}}{\partial d_{a,k}^{2}}>0$ and $\frac{\partial^{2} \mbox{r}_{k}^{*}}{\partial d_{i,k}^{2}}\frac{\partial^{2} \mbox{r}_{k}^{*}}{\partial d_{a,k}^{2}}-\Bigg(\frac{\partial^{2}\mbox{r}_{k}^{*}}{\partial d_{i,k} \partial d_{a,k}}\Bigg)^{2}>0$. Therefore, the Hessian is positive definite and $\mbox{r}_{k}^{*}$ is a convex function of $d_{i,k}$ and $d_{a,k}$.

\section{}
\label{sec:ApB}
We prove Lemma 3 by following the same reasoning of Appendix A. Note that $\bar{\boldsymbol{\mbox{r}}}_{apx}^{*}$ in~\eqref{eq:avg_rate_appx} is the sum of $K+1$ functions each of them depending only on the robot's position~$q_{k}$: 
\begin{align*} 
&\bar{\boldsymbol{\mbox{r}}}_{apx}^{*}=\frac{1}{K}\sum_{k=0}^{K}\mbox{r}_{app,k}^{*}=\stepcounter{equation}\tag{\theequation} \label{eq:rappk}\\
&\frac{1}{K}\sum_{k=0}^{K}B_{w}\mbox{log}_{2}\left(1+\Big(\widehat{A} d_{i,0,k}^{-\widehat{\nu}}+ \widehat{B}  d_{i,0,k}^{-\widehat{\nu}/2} d_{a,0,k}^{-\widehat{\mu}/2}+\widehat{C} d_{a,0,k}^{-\widehat{\mu}} \Big) \frac{p_{t}}{\sigma^{2}}\right)\\
&-\frac{\partial \mbox{r}_{k}^{*}}{\partial d_{a,k}}|_{\big(d_{a,0,k},d_{i,0,k}\big)} d_{a,0,k}-\frac{\partial \mbox{r}_{k}^{*}}{\partial d_{i,k}}|_{\big(d_{a,0,k},d_{i,0,k}\big)} d_{i,0,k}\\&+\frac{\partial \mbox{r}_{k}^{*}}{\partial d_{a,k}}|_{\big(d_{a,0,k},d_{i,0,k}\big)}\sqrt{(z_{r}-z_{a})^{2}+(x_{k}-x_{a})^{2}+(y_{k}-y_{a})^{2}}\\
&+\frac{\partial \mbox{r}_{k}^{*}}{\partial d_{i,k}}|_{\big(d_{a,0,k},d_{i,0,k}\big)}\sqrt{(z_{r}-z_{i})^{2}+(x_{k}-x_{i})^{2}+(y_{k}-y_{i})^{2}},\\
\end{align*}
where, $\frac{\partial \mbox{r}_{k}^{*}}{\partial d_{i,k}}<0$ and $\frac{\partial \mbox{r}_{k}^{*}}{\partial d_{a,k}}<0$ are given by~\eqref{eq:rate_dev1} and~\eqref{eq:rate_dev2}, respectively. Note that in~\eqref{eq:rappk} we have replaced $q_{k}$ with $[x_{k},y_{k}]$. Let us define $D_{k}=\sqrt{(z_{r}-z_{a})^{2}+(x_{k}-x_{a})^{2}+(y_{k}-y_{a})^{2}}$, then, we can compute the partial derivatives of $\mbox{r}_{app,k}^{*}$ with respect to $x_{k}$ and $y_{k}$ as follows:
\begin{align*} 
\frac{\partial\mbox{r}_{app,k}^{*}}{\partial x_{k}}=&\frac{\partial \mbox{r}_{k}^{*}}{\partial d_{a,k}}|_{\big(d_{a,0,k},d_{i,0,k}\big)}\frac{(x_{k}-x_{a})} {D_{k}}+\\&\frac{\partial \mbox{r}_{k}^{*}}{\partial d_{i,k}}|_{\big(d_{a,0,k},d_{i,0,k}\big)}\frac{(x_{k}-x_{i})} {D_{k}},\stepcounter{equation}\tag{\theequation} \label{eq:rate_app_dev1}
\end{align*}
\begin{align*} 
\frac{\partial\mbox{r}_{app,k}^{*}}{\partial y_{k}}=&\frac{\partial \mbox{r}_{k}^{*}}{\partial d_{a,k}}|_{\big(d_{a,0,k},d_{i,0,k}\big)}\frac{(y_{k}-y_{a})} {D_{k}}+\\&\frac{\partial \mbox{r}_{k}^{*}}{\partial d_{i,k}}|_{\big(d_{a,0,k},d_{i,0,k}\big)}\frac{(y_{k}-y_{i})} {D_{k}},\stepcounter{equation}\tag{\theequation} \label{eq:rate_app_dev2}
\end{align*}
The second order partial derivatives are given by:
\begin{align*} 
\frac{\partial^{2}\mbox{r}_{app,k}^{*}}{\partial x_{k}^{2}}=&\frac{\partial \mbox{r}_{k}^{*}}{\partial d_{a,k}}|_{\big(d_{a,0,k},d_{i,0,k}\big)}\frac{(y_{k}-y_{a})^{2}+(z_{k}-z_{a})^{2}} {D_{k}^{3}}+\\
&\frac{\partial \mbox{r}_{k}^{*}}{\partial d_{i,k}}|_{\big(d_{a,0,k},d_{i,0,k}\big)}\frac{(y_{k}-y_{i})^{2}+(z_{k}-z_{i})^{2}} {D_{k}^{3}},\stepcounter{equation}\tag{\theequation} \label{eq:rate_app_dev11}
\end{align*}
\begin{align*} 
\frac{\partial^{2}\mbox{r}_{app,k}^{*}}{\partial y_{k}^{2}}=&\frac{\partial \mbox{r}_{k}^{*}}{\partial d_{a,k}}|_{\big(d_{a,0,k},d_{i,0,k}\big)}\frac{(x_{k}-x_{a})^{2}+(z_{k}-z_{a})^{2}} {D_{k}^{3}}+\\
&\frac{\partial \mbox{r}_{k}^{*}}{\partial d_{i,k}}|_{\big(d_{a,0,k},d_{i,0,k}\big)}\frac{(x_{k}-x_{i})^{2}+(z_{k}-z_{i})^{2}} {D_{k}^{3}},\stepcounter{equation}\tag{\theequation} \label{eq:rate_app_dev22}
\end{align*}
\begin{align*} 
\frac{\partial^{2}\mbox{r}_{app,k}^{*}}{\partial x_{k} \partial y_{k}}=&\frac{\partial \mbox{r}_{k}^{*}}{\partial d_{a,k}}|_{\big(d_{a,0,k},d_{i,0,k}\big)}\frac{(x_{k}-x_{a})(y_{k}-y_{a})} {D_{k}^{3}}+\\
&\frac{\partial \mbox{r}_{k}^{*}}{\partial d_{i,k}}|_{\big(d_{a,0,k},d_{i,0,k}\big)}\frac{(x_{k}-x_{i})(y_{k}-y_{i})} {D_{k}^{3}},\stepcounter{equation}\tag{\theequation} \label{eq:rate_app_dev12}
\end{align*}
Note that $\frac{\partial^{2}\mbox{r}_{app,k}^{*}}{\partial x_{k}^{2}}<0$ and $\frac{\partial^{2}\mbox{r}_{app,k}^{*}}{\partial y_{k}^{2}}<0$ because $\frac{\partial \mbox{r}_{k}^{*}}{\partial d_{i,k}}|_{\big(d_{a,0,k},d_{i,0,k}\big)}<0$ and  $\frac{\partial \mbox{r}_{k}^{*}}{\partial d_{a,k}}|_{\big(d_{a,0,k},d_{i,0,k}\big)}<0$ are negative terms $\forall d_{a,0,k}>0, \forall d_{i,0,k}>0 $. Furthermore, we have that $\frac{\partial^{2}\mbox{r}_{app,k}^{*}}{\partial x_{k}^{2}}\frac{\partial^{2}\mbox{r}_{app,k}^{*}}{\partial y_{k}^{2}}-\Bigg(\frac{\partial^{2}\mbox{r}_{app,k}^{*}}{\partial x_{k} \partial y_{k}}\Bigg)^{2}>0$. 
Therefore, by follwing the leading principal minors criteria, the Hessian is negative definite and $\mbox{r}_{app,k}^{*}$ is a concave function of $q_{k}=[x_{k},y_{k}]$.

\section{}
\label{sec:ApC}
In this appendix, we prove that the sequence of solutions provided by \textit{RMAP} converges to a KKT point of $P3$ if, in each iteration $j$, $\bar{\boldsymbol{\mbox{r}}}_{map,j}\ge r_{min}$. We first observe that, when this condition holds, Lines 7-9 of \textit{RMAP} does not affect the solution. Then, solving $P3$ by \textit{RMAP}, is equivalent to solving $P3$ by using an SCO algorithm for which the convergence to a KKT point of $P3$ follows from~\cite{Conv}. The convergence is guaranteed for every feasible initial solution and every trust region size $T_{k}>0$. This proves the second part of Theorem~1.

\section{}
\label{sec:ApD}
Now, we prove that \textit{RMAP} converges even when condition $\bar{\boldsymbol{\mbox{r}}}_{map,j}\ge r_{min}$ does not hold for each iteration. More precisely, for each iteration $n$ such that $\bar{\boldsymbol{\mbox{r}}}_{map,n}< r_{min}$, \textit{RMAP} sets $\boldsymbol{q}_{n}=\boldsymbol{q}_{n-1}$ and decreases $T_{k}$ for a certain position $q_{k}$. Then, we can have one of the following three cases:
\begin{itemize}
\item \textit{Case 1}: $T_{k}>0, \; \forall k$. \textit{RMAP} continues solving problem $P3$ from iteration $j=n+1$ with $\boldsymbol{q}_{n-1}$ as the initial feasible solution and $T_{k}> 0, \; \forall k$.  As proved in Appendix C, if for the successive iterations, i.e., $\forall j>n+1$, we have that $\bar{\boldsymbol{\mbox{r}}}_{map,j}\ge r_{min}$, \textit{RMAP} still converges to a KKT point of $P3$. Otherwise, for each iteration $n$ such that $\bar{\boldsymbol{\mbox{r}}}_{map,n}< r_{min}$, Lines 8 and 9 are repeated and this proof follows either \textit{Case 1} or \textit{Case 2} whether $T_{k}>0, \; \forall k$ or $T_{k}=0$ for a certain position, respectively. If $T_{k}=0, \; \forall k$, the proof follows \textit{Case 3}.
\item \textit{Case 2}: $T_{k}=0$ for a certain position~$q_{k}$ and iteration $n$. In this case, we have that constraint~\eqref{Con_Trust} of $P4$ becomes: 
\begin{align*} 
\lVert q_{j,k}-q_{j-1,k} \lVert_{2}  \le T_{k}=0, \;\forall j>n.
\end{align*}
This is equivalent to adding the following affine constraints to $P3$ and $P4$, respectively:
\begin{align*} 
q_{k}=q_{n-1,k} \;\; and \;\; q_{j,k}=q_{n-1,k}, \forall j>n.
\end{align*}
The constraints, in fact, fix the position of the robot~$q_{k}$. Given these constraints, we are free to set $T_{k}>0$. Thus, from iteration $j>n$, \textit{RMAP} solves a modified version of $P3$ ($P3'$), starting from initial solution $q_{n-1}$ and $T_{k}>0, \forall k$ by iteratively solving a modified version of $P4$ ($P4'$). Note that adding this affine constraint to $P3$ and $P4$ does not change the convexity of the latter. As done in Appendix C, we can prove that \textit{RMAP} converges to a KKT point of $P3'$ if $\forall j>n+1$, we have that $\bar{\boldsymbol{\mbox{r}}}_{map,j}\ge r_{min}$ holds. Otherwise, if $\bar{\boldsymbol{\mbox{r}}}_{map,j}< r_{min}$ for a certain iteration, the proof follows \textit{Case 1} or \textit{Case 2}, whether 
$T_{k}>0, \; \forall k$ or $T_{k}=0$ for a certain position, respectively. If $T_{k}=0, \; \forall k$, the proof follows \textit{Case 3}.
\item \textit{Case 3}: $T_{k}=0, \; \forall k$. In this case, the robot's position for each $k=0,...,K$ is fixed, and \textit{RMAP} has converged to a solution.
\end{itemize}
Thus, we have that \textit{RMAP} either converges to a KKT point of $P3$ or to a KKT point of a modified problem where, for some or for all $k=0,...,K$, the robot's positions are fixed.

\bibliography{ref}
\bibliographystyle{IEEEtran}

\end{document}